\documentclass[journal,compsoc,screen]{IEEEtran}

\usepackage{hyperref}
\usepackage{mdframed}
\usepackage{lipsum}
\usepackage{amsmath}
\usepackage{amsfonts}
\usepackage[font=small]{caption}
\usepackage{ifpdf}
\usepackage{todonotes}
\usepackage{booktabs} 
\usepackage{url}
\usepackage{pifont}
\usepackage{color}
\usepackage{tcolorbox}
\usepackage{balance}
\usepackage{makecell}
\usepackage{verbatim}
\usepackage{tabularx}
\usepackage{colortbl}

\usepackage{algorithmic}
\usepackage{array}

\usepackage{bussproofs}
\usepackage{lipsum} 
\usepackage{mathpartir}

\usepackage{caption}
\usepackage{multirow}
\usepackage{float}
\usepackage{mwe}
\newcommand{\code}[1]{\texttt{#1}}

\usepackage{url}
\usepackage{todonotes}
\usepackage{subfigure}
\usepackage{tikz}
\usepackage{xcolor}

\usepackage{booktabs,caption,fixltx2e}
\usepackage[flushleft]{threeparttable}

\hypersetup{
 colorlinks=true,
 linkcolor=blue,
 citecolor=blue,
 filecolor=black,
 urlcolor=black,
 pdfauthor = {},
 pdftitle = {},
 pdfkeywords = {}
}

\usepackage{listings}
\usepackage{algorithm2e}
\usepackage{xcolor}

\SetAlFnt{\footnotesize}

\SetAlCapSty{xAlCapSty}

\SetCommentSty{xCommentSty}

\SetNlSty{mynlfont}{}{} 

\LinesNumbered

\SetSideCommentRight

\DontPrintSemicolon

\RestyleAlgo{algoruled}

\begin{document}

\title{ChatGPT vs SBST: A Comparative Assessment of Unit Test Suite Generation}

\author{Yutian~Tang, 
Zhijie~Liu, 
Zhichao~Zhou, and
Xiapu~Luo
\thanks{Yutian Tang is with University of Glasgow, United Kingdom. E-mail: csytang@ieee.org.}
\thanks{Zhijie Liu is with ShanghaiTech University, Shanghai 201210, China. E-mail: liuzhj2022@shanghaitech.edu.cn.}
\thanks{Zhichao Zhou is with ShanghaiTech University, Shanghai 201210, China. E-mail: zhouzhch@shanghaitech.edu.cn.}
\thanks{Xiapu Luo is with the Department of Computing, Hong Kong Polytechnic University, Hong Kong SAR, China. E-mail: csxluo@comp.polyu.edu.hk.}
\thanks{Yutian Tang (csytang@ieee.org) is the corresponding author.}}

\markboth{Journal of \LaTeX\ Class Files,~Vol.~14, No.~8, August~2021}%
{Shell \MakeLowercase{\textit{et al.}}: A Sample Article Using IEEEtran.cls for IEEE Journals}


\IEEEtitleabstractindextext{

\begin{abstract}
    Recent advancements in large language models (LLMs) have demonstrated exceptional success in a wide range of general domain tasks, such as question answering and following instructions. Moreover, LLMs have shown potential in various software engineering applications. In this study, we present a systematic comparison of test suites generated by the ChatGPT LLM and the state-of-the-art SBST tool EvoSuite. Our comparison is based on several critical factors, including correctness, readability, code coverage, and bug detection capability. By highlighting the strengths and weaknesses of LLMs (specifically ChatGPT) in generating unit test cases compared to EvoSuite, this work provides valuable insights into the performance of LLMs in solving software engineering problems. Overall, our findings underscore the potential of LLMs in software engineering and pave the way for further research in this area.
\end{abstract}


}

\maketitle

\section{Introduction}

Unit testing is a widely accepted approach to software testing that aims to validate the functionality of individual units within an application. By using unit tests, developers can detect bugs in the code during the early stages of the software development life cycle and prevent changes to the code from breaking existing functionalities, known as regression \cite{Zhu:1997}. The primary objective of unit testing is to confirm that each unit of the software application performs as intended. This method of testing helps to improve the quality and reliability of software by identifying and resolving issues early on.

\noindent\textbf{SBST.} The importance of unit testing in software development and the software development life cycle cannot be overstated. To generate unit test cases, search-based software testing (SBST) \cite{Mark:12} techniques are widely employed. SBST is a technique that employs search algorithms such as genetic algorithms and simulated annealing to create test cases. The objective of SBST is to utilize these kinds of algorithms to optimize the test suites, resulting in a set of test cases that provide extensive code coverage and effective detection of program defects. Compared to other testing techniques, SBST exhibits promising results in reducing the number of test cases while maintaining the same level of defect detection capability \cite{Fraser:13, Zhou:23}. SBST has emerged as an effective approach to improving the quality and efficiency of software testing, providing a valuable tool for software developers to streamline the testing process.

\noindent\textbf{Large Language Model and ChatGPT.} Recently, Large language models (LLMs) have exhibited remarkable proficiency in processing and performing everyday tasks such as machine translation, question answering, summarization, and text generation with impressive accuracy \cite{Carlini:21,Brants:07,Raffel:22}. These models possess nearly the same capacity as humans for understanding and generating human-like text. One such example of a real-world LLM application is OpenAI's GPT-3 (Generative Pretrained Transformer 3), which has been trained on an extensive amount of text data from the internet. Its practical implementation, ChatGPT \footnote{CharGPT: The version used in this study is GPT-3 instead of GPT-4}, is widely employed in various daily activities, including text generation, language translation, question answering, and automated customer support. ChatGPT has become an essential tool for many individuals, simplifying various tasks and improving overall efficiency.

\noindent\textbf{Deep-learning based Test Case Generation.} Besides accomplishing daily tasks, such as text generation, language translation, and question answering, large language models are also been adopted and used to cope with software engineering (SE) tasks, such as, code generation \cite{Svyatkovskiy:20,Alon:20,Poesia:22}, code summarization \cite{McBurney:16,Haiduc:10,Zhang:20ICSE}, document and comments generation \cite{McBurney:14,Hu:18}, and more.  These models can be employed to generate unit test cases for programs with the help of a large number of real-world test cases written by developers/testers. This allows for the validation of the intended functionality of individual units within the software application. The integration of LLMs in SE tasks has demonstrated their versatility and potential for improving software development processes.

\noindent\textbf{Motivation.} Despite the SBST performing well in generating unit tests, there is still a learning cost for test personnel with limited experience. As a result, it can be a barrier to embracing SBST techniques, especially for fresh testers. However, the applications based on large language models can accomplish the same task (i.e., generating test suites) with nearly no learning costs. However, it is still unknown whether the unit tests generated by SBST can be compared with advanced artificial intelligence models and techniques. For example, whether the LLM-generated test cases are readable, understandable, reliable, and can be used in practice. Here, in this paper, we are interested in understanding the strengths and weaknesses of test suites generated by LLM. Specifically, we leverage the state-of-art GPT-3 \cite{Brown:20} model's product ChatGTP \cite{ChatGPT,Brown:20} as a representative of LLM for comparison. More importantly, this paper intends to gain insights from two aspects: (1) we are keen on the knowledge we can learn from large language models to improve the state-of-art SBST techniques, and (2) we are also interested in uncovering the potential limitations of the existing large language models in generating test suite. 


\noindent\textbf{Our Study.} To cope with the aforementioned challenges and achieve the goals, in this paper, we intend to answer the following research questions (RQ):

\noindent$\bullet$ \textbf{RQ1 (Correctness):} Are ChatGPT’s unit test suite suggestions correct? 

\noindent$\bullet$ \textbf{RQ2 (Readability):} How understandable is the test suite provided by ChatGPT?

\noindent$\bullet$ \textbf{RQ3 (Code Coverage):} How does ChatGPT perform with SBST in terms of code coverage?

\noindent$\bullet$ \textbf{RQ4 (Bug Detection):} How effective are ChatGPT and SBST in generating test suites that detect bugs?

\noindent\textbf{Contribution.} In summary, we make the following contributions in this paper:

\noindent$\bullet$ In this paper, we conduct the \emph{first} comparative assessment of LLMs and SBST in terms of generating unit test suites for programs in Java programming language;

\noindent$\bullet$ We systematically evaluate the test suites generated by ChatGPT from various aspects, including correctness, readability, code coverage, bug detection capability; and 

\noindent$\bullet$ Our findings contribute to a better understanding of the potential for LLMs to improve software engineering practices, specifically in the domain of unit test generation.

\section{Background}

\noindent\textbf{SBST and Evosuite.}
Search-based software testing (SBST) is a technique that formulates unit test generation as the optimization problem \cite{Tonella:04}. SBST regards code coverage as the test generation's target (e.g., branch coverage) and describes it as a fitness function to guide genetic algorithms \cite{Fraser:13,Panichella:15,Panichella:18}. The genetic algorithms evolve tests by iterating to (1) apply mutation and crossover operators to existing tests (i.e., the current generation) for new offspring tests and (2) form a new generation by selecting those with better fitness scores from the current generation and offspring. In our work, we choose the most mature SBST tool in Java, Evosuite \cite{Fraser:11}.

\begin{figure}[!htpb]
	\centering
	\includegraphics[width=0.5\textwidth] {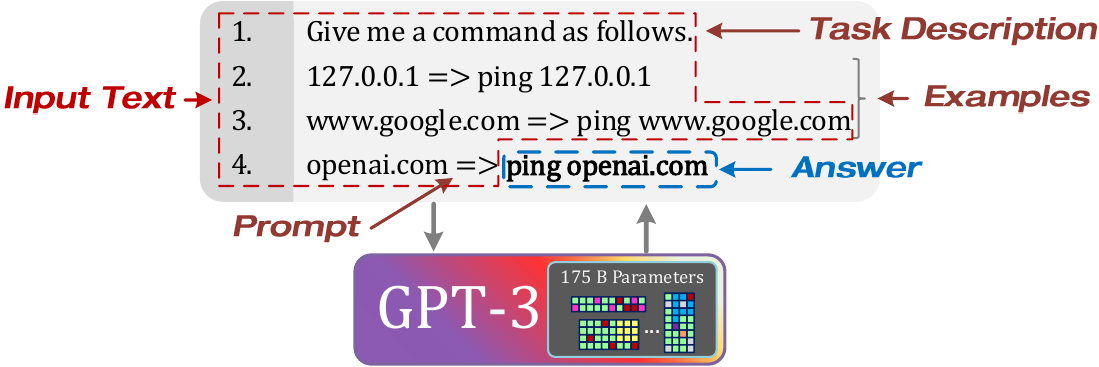}
	\caption{A Sample Use of GPT-3}
    \label{fig:sample-use-GPT-3}
\end{figure}

\noindent\textbf{LLM and ChatGPT.} 
LLM is the type of biggest model in terms of parameter count, trained on enormous amounts of text data (e.g., human-like text, code, and so on)~\cite{Vaswani:17, Devlin:18, Raffel:20, Brown:20, Ouyang:22, ChatGPT}. It is designed to process and understand input natural language text and to generate text consistent with the input, and shows a strong ability in natural language processing (NLP) tasks, such as, machine translation, question answering, text generation, and so on. ChatGPT~\cite{ChatGPT} is now the most ideal LLM (i.e., adapt to human expression by using Instruct)~\cite{Artetxe:22, Ouyang:22} implemented atop GPT-3. GPT-3~\cite{Brown:20} is constructed on multi-layer Transformer decoders~\cite{Vaswani:17, radford:19, radford:18} with 175 billion parameters, using few-shot learning (i.e., multiple examples and prompt). It shows performance similar to that of state-of-art fine-tuned systems in many tasks. One example of using GPT-3 is shown in Fig. \ref{fig:sample-use-GPT-3}. GPT-3 takes in the input text and infers the answer based on the task description, examples, and prompts in the input. To make LLM further align with users (humans), InstructGPT~\cite{Ouyang:22} utilizes additional supervised learning and reinforcement learning from human feedback to fine-tune GPT-3. ChatGPT~\cite{ChatGPT} uses the same methods as InstructGPT and has the ability to answer follow-up questions. 

For generating unit test cases, one can utilize a large language model like GPT-3. To generate new test cases given code snippets as input, the model can be fine-tuned on a dataset of code snippets and their accompanying test cases. One can also take advantage of ChatGPT's answering follow-up questions to generate more diverse test suites for given code snippets.

\begin{figure}[!htpb]
	\centering
	\includegraphics[width=0.5\textwidth] {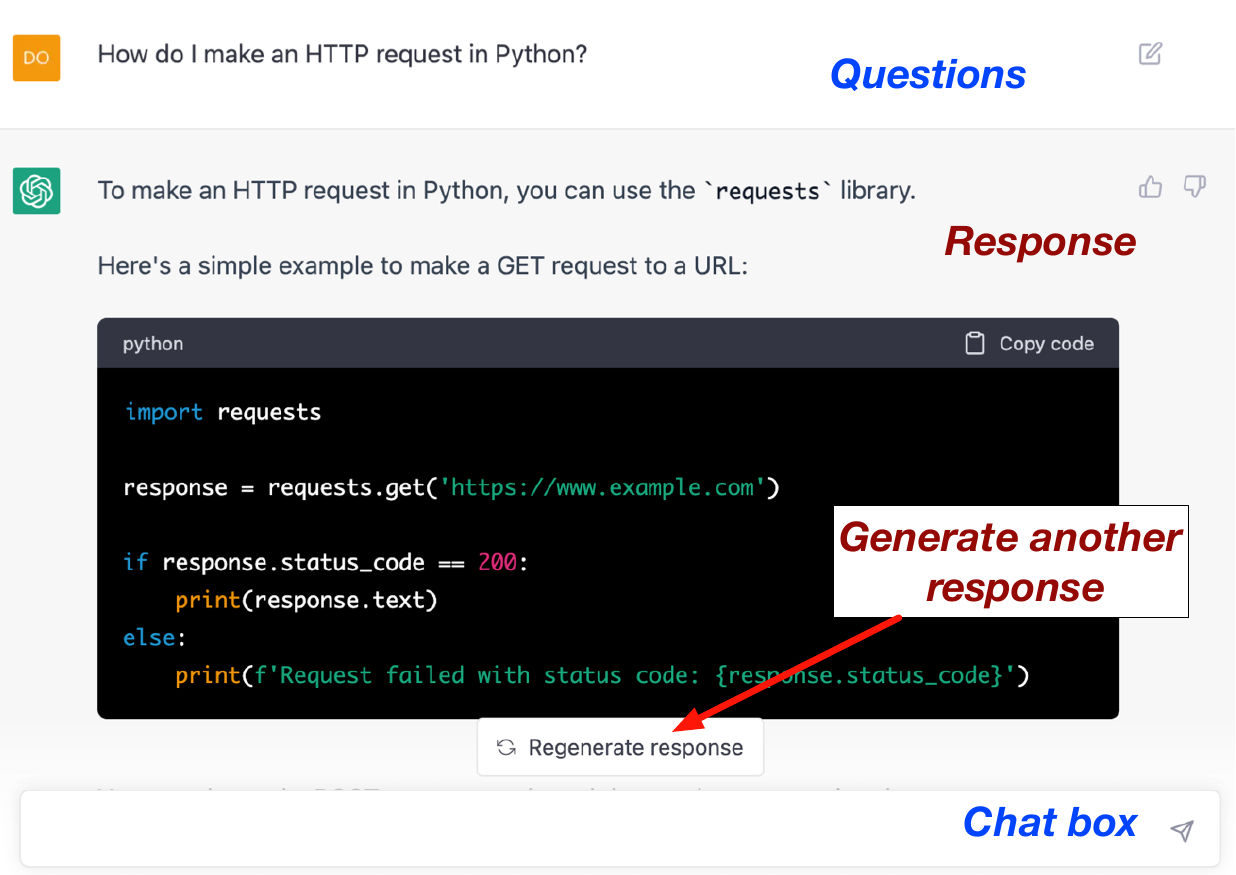}
    \vspace{-1em}
	\caption{A Sample Use of ChatGPT in a SE Task}
 \vspace{-1em}
    \label{fig:sample-use-ChatGPT}
\end{figure}

\noindent\textbf{Using of ChatGPT.} ChatGPT \cite{ChatGPT,Brown:20} can be used as follows. The software developer/tester (user) registers an account for ChatGPT. Then, users send a prompt (a text or a question) to ChatGPT. Then, ChatGPT will respond based on the information it has learned from its training data. Also, ChatGPT can be used in most software-engineering related tasks, such as, generating code, generating comments, and generating test cases. For example, as shown in Fig. \ref{fig:sample-use-ChatGPT}, ChatGPT offers a basic user interface like a Chatbot, in which a user can ask any question in a natural language. As shown in Fig. \ref{fig:sample-use-ChatGPT}, we ask ChatGPT how to make an HTTP request in Python, and ChatGPT shows a sample code written in Python with corresponding explanations. If a user is not satisfied with the generated responses, (s)he can ask ChatGPT to regenerate a response by clicking the ``Regenerate a response'' button at the bottom of the page. 

\section{Comparative Assessment Setup}

\subsection{Data Collection}
As for \textbf{RQ1-3}, to reduce bias in selecting subject code for generating test cases, we reuse the existing benchmark used in the existing study to evaluate the performance of Evosuite. Here, we use the benchmark presented in DynaMOSA (a.k.a Dynamic Many-Objective Sorting Algorithm) \cite{Panichella:18}. The benchmark contains 346 Java classes from 117 projects. The detailed class information can be founded in \cite{Panichella:18} and our artifact repository (Sec.\ref{sec:data}). However, based on facts reported by other works \cite{Zhou:23,Lin:21}, some projects in the SF100 dataset can be obsolete and
are no longer maintained. Some projects are not able to build and compile as some classes required in DynaMOSA dataset are missing or not publicly available. As a result, we remove 38 projects and remain 79 projects with 248 Java classes. As for \textbf{RQ4}, we use the state-of-art defect database for Java-related research, which is Defects4J \cite{Defects4J}. It contains 835 bugs from 17 open-source projects. 

\subsection{Using ChatGPT to Generate Unit Test Cases}\label{subsec:chatgpt_generate_test_case}

With the help of ChatGPT, we are able to automatically generate unit test cases for programs. Unfortunately, there is no standard or oracle on how to use ChatGPT to automatically generate unit test cases with ChatGPT. Therefore, we adopt the following step to learn a reasonable practice of using ChatGPT to generate unit test cases:

\noindent$\bullet$ \textbf{Step 1.} Collecting existing tools that leverage LLM (e.g., ChatGPT) to automatically generate unit test cases from various sources, including Google, Google Scholar, GitHub, and technical blogs;

\noindent$\bullet$ \textbf{Step 2.} Analyzing the phrases and descriptions used in these tools to prompt LLM to generate test cases. This part involves analyzing source codes, reading blocks, and learning technical documents; and

\noindent$\bullet$ \textbf{Step 3.} Verifying the phrases and descriptions collected in Step 2 with ChatGPT to exclude invalid phrases and descriptions;

Through the \textbf{Step 1-3}, we obtain the following representative expressions that are able to generate unit test cases for a code segment:

\noindent$\bullet$ \textbf{Expression 1:} ``Write a unit test for \$\{input\}'' with the code segment under test as the input;

\noindent$\bullet$ \textbf{Expression 2:} ``Can you create unit tests using JUnit for \$\{input\}?'' with the code segment under test as the input;

\noindent$\bullet$ \textbf{Expression 3:} ``Create a full test with test cases for the following Java code: \$\{input\}?'' with the code segment under test as the input; 

Based on the above findings, we summarize our prompt as: ``Write a JUnit test case to cover methods in the following code (one test case for each method): \$\{input\}?'' with the code segment under test as the input. Note that, to mimic the real-world practice, we do not intend to compare and evaluate the ChatGPT prompts to build a best-performed prompt. Instead, we only intend to \textbf{build a reasonable prompt} for ChatGPT to stimulate how developers use ChatGPT in a real-world environment.

\subsection{Other Setups for the Study}\label{subsec:other_setups}

\noindent$\bullet$ \textbf{Setup for EvoSuite.} EvoSuite provides many parameters
(e.g., crossover probability, population size \cite{Evosuite}) to run the algorithms. In this paper, to evaluate and compare the performance between Evosuite and ChatGPT, we remain the default settings in Evosuite. As Evosuite leverages genetic algorithms in selecting and generating test cases, to reduce the bias introduced by randomness, we run 30 times for each class.

\noindent$\bullet$ \textbf{Long Inputs for ChatGPT.}
The maximum input length for ChatGPT is 2,048 tokens, which is roughly equivalent to 340-350 words. If the input submitted is too long, ChatGPT reports an error message and gives no response. In this case, we try to split the entire class by methods and ask ChatGPT to generate unit test cases for methods. However, splitting the entire class by methods to generate test cases cannot be a good practice as some information about the entire class cannot be perceived by ChatGPT. As a result, it hurts the quality of generated test cases. Here, we set the maximum length to be 4,096 tokens. That is, if the length of a class is larger than 4,096 tokens, we discard it.

\noindent$\bullet$ \textbf{Environment.} Experiments on EvoSuite are conducted on a machine with Intel(R) Core(TM) i9-10900 CPU @ 2.80GHz and 128 GB RAM.

\section{Experiment and Evaluation}\label{sec:experiments}

\subsection{Correctness}

\noindent\textbf{RQ1: Are ChatGPT’s Unit Test Suite Suggestions Correct?}

\noindent\textbf{Motivation.} The first and foremost thing we need to examine is whether ChatGPT can correctly return the test cases for testing the program/code segment given.

\noindent\textbf{Methodology.} To test whether the generated test cases are correct. We need to evaluate them from three aspects: (1) whether ChatGPT successfully returns the test case for each input under test; (2) whether these test cases can be compiled and executed; and (3) whether these test cases contain potential bugs. Specifically, for (2), it can be examined with the help of Java Virtual Machine (JVM). We compile and execute the test cases to see whether JVM reports errors. For (3), we rely on the state-of-art static analyzer, SpotBugs \cite{SpotBugs,Findbugs,Ayewah:08}, to scan the test cases generated by ChatGPT to find out whether these test cases contain potential bugs or vulnerabilities. SpotBugs \cite{SpotBugs} is the successor of FindBugs \cite{Findbugs,Ayewah:08} (an abandoned project) and is an open-source static software analyzer, which can be used to capture bugs in a Java program. It supports more than 400 bug patterns and poor programming practices. 

\noindent\textbf{Results.}
According to the \textit{Long-input setting} in Sec. \ref{subsec:other_setups}, we remove 41 classes and remain 207 Java classes from 75 projects.



We find that ChatGPT can successfully generate unit test cases for all 207 Java classes without reporting any errors. Among these test cases, there are 144 (69.6\%) test cases can be successfully compiled and executed without needing extra-human efforts. Next, we ask two undergraduate students who have basic knowledge of Java programming to attempt to repair errors with the help of IntelliJ IDE \cite{IntelliJ}. For the rest 64 test cases, there are 3 test cases that cannot be directly fixed without the background knowledge of the target program, and 60 test cases can be repaired with the help of IDE. Specifically, the errors in 3 test cases fall into 3 categories: a) fail to implement an interface; b) fail to initiate an abstract class instance; c) try to initiate an instance of an inner class. 

\begin{table}[!htpb]
\centering
\caption{Error Types in 60 Test Cases}
\begin{tabular}{cc}
\hline
\textbf{Type of Errors} & \textbf{Frequency} \\ \hline
\textbf{Access Private/Protected Field} & 31 \\
\textbf{Access Private/protected Methods} & 20 \\
\textbf{Invoke undefined methods} &  11  \\
\textbf{Fail to initiate an instance for an interface} & 10   \\
\textbf{Incorrect parameter type} & 2\\
\textbf{Fail to initiate an instance} & 2 \\
\textbf{Access undefined field} & 1  \\ \hline
\end{tabular}
\label{tab:type-of-errors}
\end{table}

The errors in other 60 test cases fall into 7 categories as shown in Table. \ref{tab:type-of-errors}. Here, \textit{invoke undefined methods} represents invoking a method, which is not defined in the target class. Table. \ref{tab:invoke-undefined-methods} shows some samples of invoking undefined method errors. The root cause for invoking undefined methods is that ChatGPT is only given the class under test instead of the entire project. As a result, ChatGPT has to predict the name of a callee when needed. This is especially the case when ChatGPT attempts to generate some \textit{Assertions}. However, the results in Table. \ref{tab:invoke-undefined-methods} also surprise us that even if the ChatGPT fails to call the correct callees, its prediction also gives a strong clue to find the correct callee names. This is why we can fix these errors without the need of domain knowledge of these target projects. \textit{Fail to initiate an instance for an interface} represents that ChatGPT creates an instance of an interface, but fails to override methods, and \textit{incorrect types} represents that the types of arguments in callsites are incorrect.

\begin{table}[!htpb]
\centering
\caption{Examples of Invoking Undefined Methods}
\scalebox{0.78}{
\begin{tabular}{cccc}
\hline
\textbf{Project} & \textbf{Classes} & \textbf{ChatGPT's CallSite} & \textbf{Correct CallSite} \\ \hline
trove & TFloatDoubleHash & hash.\textcolor{red}{get(val)} & hash.\textcolor[RGB]{0,128,28}{index(val)} \\
trove & TFloatDoubleHash & hash.\textcolor{red}{put(3, 4.0f)} & hash.\textcolor[RGB]{0,128,28}{insertKeyAt(3, 4.0f)}  \\
24\_saxpath & XPathLexer & token.\textcolor{red}{getStart()} & token.\textcolor[RGB]{0,128,28}{getTokenBegin()}\\
24\_saxpath & XPathLexer & token.\textcolor{red}{getType()} & token.\textcolor[RGB]{0,128,28}{getTokenType()} \\
73\_fim1 & UpdateUserPanel & user.\textcolor{red}{setUsername} & user.\textcolor[RGB]{0,128,28}{setName()} \\ \hline
\end{tabular}
}
\label{tab:invoke-undefined-methods}
\end{table}

To wrap up, the compiling errors made by ChatGPT are mainly due to that it fails to have an overview of the entire project. Thus, ChatGPT attempts to predict the callees' names, parameters, parameters' types, and so forth. As a result, compiling errors are introduced.

\noindent$\triangleright$ For (3), we leverage the state-of-art static analyzer, SpotBugs, to scan the test cases generated by ChatGPT. As a result, SpotBugs report 403 potential bugs from 204 test cases (3 test cases fail to compile). The overview distribution is shown in Table. \ref{tab:bugoverview}. On average, each case contains 1.97 bugs.

\begin{table}[!htpb]
\centering
\caption{Bug Pattern Overview}
\begin{tabular}{cc}
\hline
\textbf{Num. of Potential Bugs} & \textbf{Num. of Class} \\ \hline
Over 20 & 3 (1.47\%)\\
10 - 20 & 7 (3.43\%)\\
1 - 9 & 69 (33.8\%)\\ 
0 & 125 (61.2\%) \\ \hline
\end{tabular}
\label{tab:bugoverview}
\end{table}

\begin{table}[!htpb]
\centering
\vspace{-1em}
\caption{Bug Patterns' Priority Levels}
\vspace{-1em}
\begin{tabular}{cccc}
\hline
\textbf{Priority Level} & \textbf{\# Bugs} & \textbf{\# Related Test Cases} & \textbf{Average} \\ \hline
Scariest & \textcolor[RGB]{0,128,28}{15} & 8 (3.9\%) & \textcolor[RGB]{0,128,28}{1.87} \\
Scary & 35 & 12 (5.8\%) & 2.91\\
Troubling & 10 & 7 (3.4\%) & 1.42\\ 
Of Concern & \textcolor[RGB]{0,128,28}{343} & 70 (34.3\%) & \textcolor[RGB]{0,128,28}{4.9}\\ \hline
\end{tabular}
\label{tab:bugpattern-priority}
\end{table}

\begin{table}[!htpb]
\centering
\vspace{-1em}
\caption{Bug Patterns}
\vspace{-1em}
\scalebox{0.9}{
\begin{tabular}{cccc}
\hline
\textbf{Bug Patterns} & \textbf{\# Bugs} & \textbf{\# Related Test Cases} & \textbf{Average} \\ \hline
Bad Practice & 65 & 20 (9.8\%) & 3.25 \\
Performance & 36 & 19 (9.4\%) & 1.89 \\
Correctness & \textcolor[RGB]{0,128,28}{52} & \textcolor[RGB]{0,128,28}{20 (9.8\%)} & 2.6 \\ 
Multi-thread  Correctness & \textcolor[RGB]{0,128,28}{1} & \textcolor[RGB]{0,128,28}{1 (0.49\%} & 1 \\ 
Dodgy Code & 199 & 45 (22.2\%) &  4.42 \\ 
 Internationalization & 47 & 10 (4.9\%) & 4.7 \\ 
Experimental & 3 & 2 (0.98\%) & 1.5 \\ \hline
\end{tabular}
}
\label{tab:bugpattern}
\end{table}

From the \textbf{bug priority levels} perspective, SpotBugs rank bugs' priority level into \textit{Scariest}, \textit{Scary}, \textit{Troubling}, and \textit{Of Concern}. \textit{Scariest} level represents bugs are considered the most severe and potentially harmful to the overall functionality and security of the code; \textit{Scary} level represents bugs are considered significant and could lead to issues if not fixed; \textit{Troubling} level represents bugs are categorized as minor but could still cause issues if left unaddressed; and \textit{Of Concern} level represents bugs are considered informational and generally pose minimal to no risk to the code's functionality or security. As shown in Table. \ref{tab:bugpattern-priority}, most bugs (85.11\%) are with the \textit{Of Concern} type. There are only 8 test cases (3.9\%) that have \textit{Scariest} level bugs.

From the \textbf{bug patterns} perspective, founded bugs fall into 7 categories: (1) Bad Practice; (2) Performance; (3) Correctness; (4)Multi-thread Correctness; (5) Dodgy Code; (6) Internationalization; and (7) Experimental. The detailed descriptions of each bug pattern can be found on the official documentation \cite{SpotBugsDesc}. As shown in Table. \ref{tab:bugpattern}, there are 21 test cases involved either in correctness bugs or multi-thread correctness bugs. These types of bugs represent appear coding mistakes, which normally belong to the Scariest or Scary priority level. As for Dodgy code pattern, which holds the largest proportion, it represents the code is confusing, anomalous, or written in a way that leads itself to errors. Example cases can be dead local stores, switch fall through, and unconfirmed casts. As for \textit{correctness/multi-thread correctness} bugs, it mostly refers to the following 3 cases based on our results: null dereference, out-of-bounds array access, and unused variables.

In summary, from the bug priority levels and bug patterns, we can conclude that most (61.2\%) ChatGPT-generated test cases are bug-free. Only 20 (9.8\%) test cases are from the Scariest and Scary levels.

\begin{tcolorbox}[boxrule=1pt,boxsep=1pt,left=2pt,right=2pt,top=2pt,bottom=2pt,title=Answer to RQ1: Correctness]

\noindent$\bullet$ Of the 207 Java test cases generated, 69.6\% were compiled and executed without human intervention. However, 3 test cases were unfixable without understanding the target program and 60 could be fixed with an IDE.

\noindent$\bullet$ After analyzing the bug priority levels and bug patterns of ChatGPT-generated test cases, it can be inferred that a majority of these cases, specifically 61.2\%, are free from any bugs. However, a small proportion of test cases, comprising only 9.8\%, have been categorized under the Scariest and Scary levels, indicating the presence of severe issues. 

\end{tcolorbox} 

\subsection{Readability}

\noindent\textbf{RQ2: How Understandable is the Test Suite provided by ChatGPT?}

\noindent\textbf{Motivation.} Analyzing the readability of ChatGPT-generated code is to make sure that human developers can easily maintain, comprehend, and modify it. This is crucial when ChatGPT-generated code will be maintained and changed over time by other developers or when it will be merged into already-existing codebases.

\noindent\textbf{Methodology.} For this RQ, we set up two sub-tasks: (1) code style checking; and (2) code understandability.

\noindent$\triangleright$ To check code styles of generated test cases, we rely on the state-of-art software quality tool which supports Java: Checkstyle \cite{CheckStyle}, which is a development tool to check whether Java code adheres to a coding standard.  It automates the process of checking Java code. Here, we leverage two standards (i.e., Sun Code Conventions \cite{SUNJavaStyle}, Google Java Style \cite{GoogleJavaStyle}) with Checkstyle to check whether the ChatGPT generated test suite adheres to these standards. 

\noindent$\triangleright$ Dantas et al. \cite{Dantas:21} proposed cognitive complexity and cyclomatic complexity metrics for measuring the understandability of a code snippet. Cyclomatic complexity measures program complexity by counting independent paths in source code. It indicates code size, structure, and complexity, and helps find error-prone areas. Cognitive complexity is a metric that evaluates code complexity from a human perspective. It considers factors like code structure, naming, and indentation to determine how hard code is to understand. It helps developers gauge maintainability and modification difficulty and identifies complex or confusing code parts. Cyclomatic and cognitive complexity can be measured with the PMD IntelliJ plugin \cite{PMD}. The details can be found on the project repository (Sec. \ref{sec:data}).

\noindent\textbf{Results.} According to the \textit{Long-input setting} in Sec. \ref{subsec:other_setups}, we remove 41 classes and remain 204 Java classes from 75 projects. 

\noindent$\bullet$ \textbf{Code Style Checking Results.}

\begin{figure}[!htpb]
	\centering
	\includegraphics[width=0.5\textwidth] {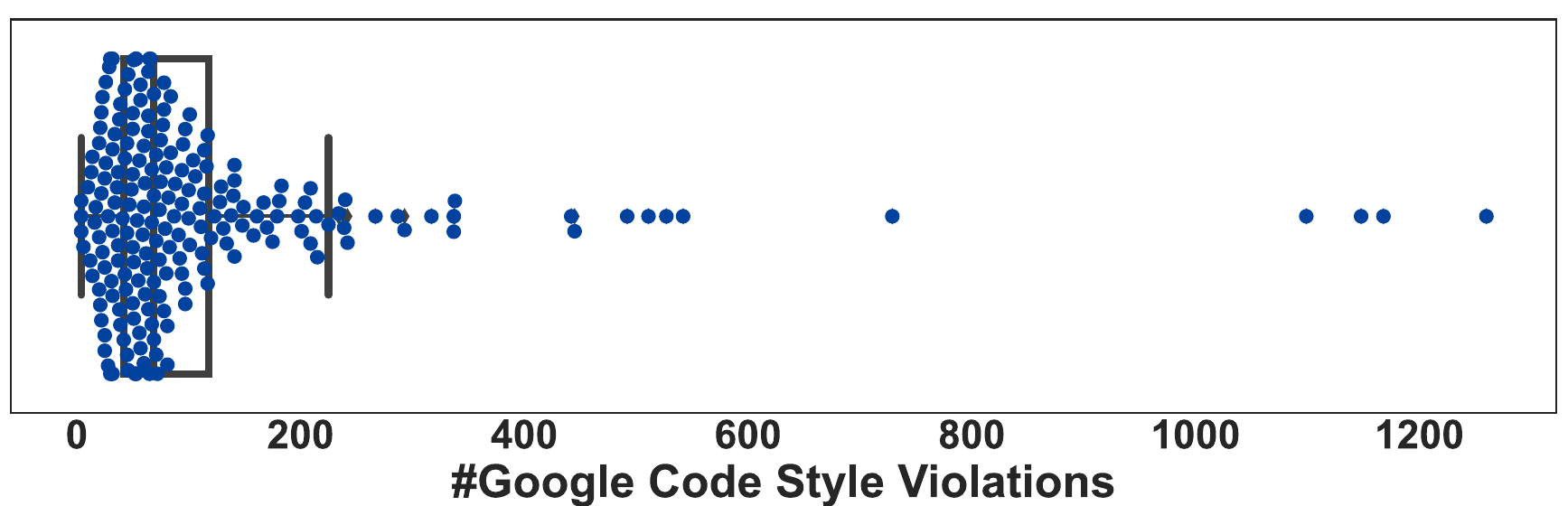}
	\caption{Boxplot of Google Code Style Violations}
    \label{fig:google-style}
\end{figure}

\noindent$\triangleright$ \underline{Checkstyle-Google:} Fig. \ref{fig:google-style} shows the boxplot of Google Codestyle violations for each class. It shows that the dataset has several outliers on the higher side, with a median value of approximately 70. The interquartile range (IQR) falls between around 30 to 175, indicating that most of the data lie within this range. However, the data is highly skewed to the right, with a few extreme data points on the higher side, indicating that the distribution is not normal. The minimum value is 4, and the maximum value is 1260, which shows a wide range of values in the dataset.

\begin{figure}[!htpb]
	\centering
	\includegraphics[width=0.4\textwidth] {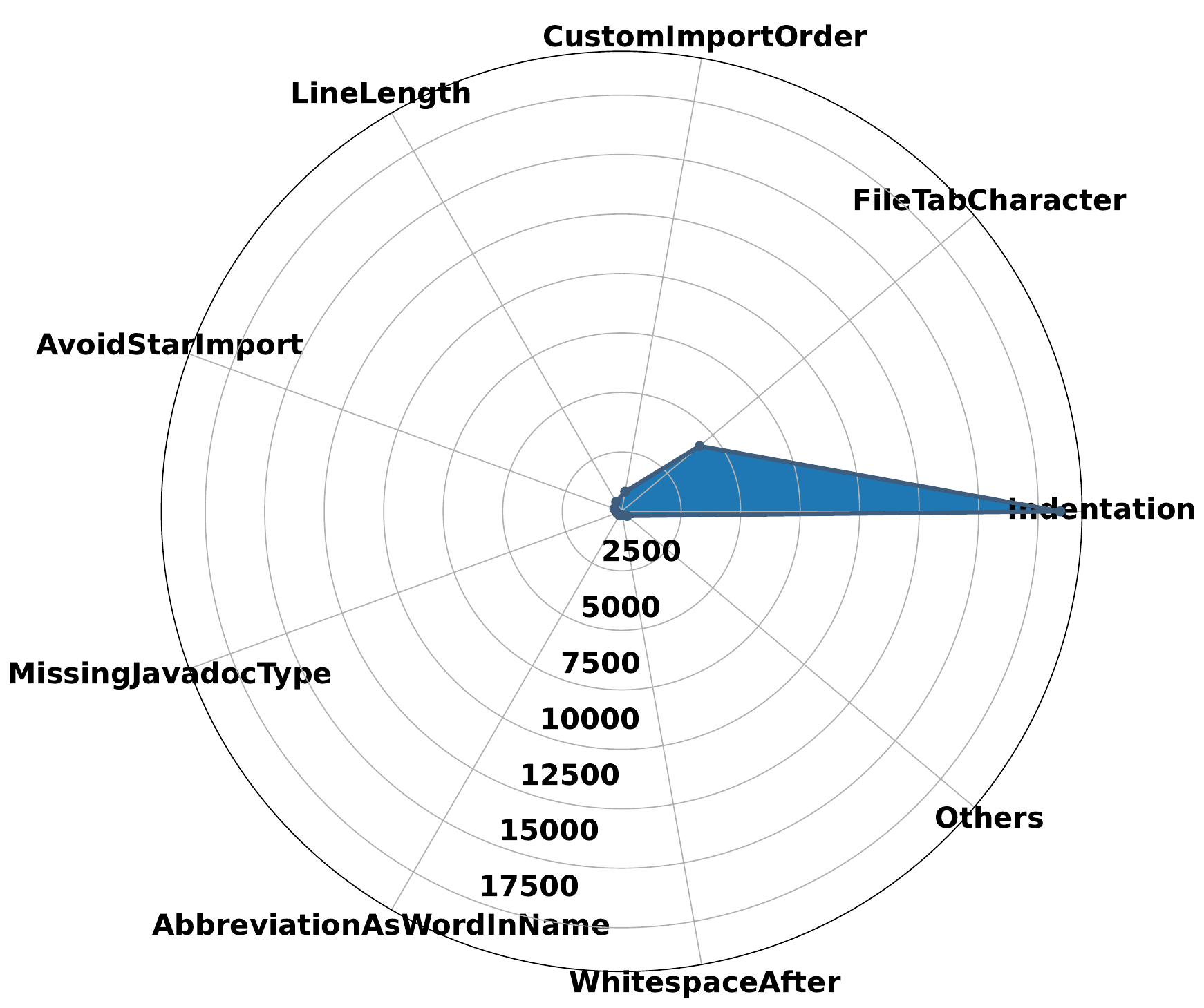}
	\caption{Radar Plot of Google Code Style Violations}
    \label{fig:google-style-radar}
\end{figure}

Next, The radar plot in Fig. \ref{fig:google-style-radar} breakdowns violation issues by types to display the details. As depicted in Fig. \ref{fig:google-style-radar}, we can conclude that:

\noindent$\bullet$ \code{Indentation} is the most common code style violation, indicating that ChatGPT may need to work on consistently formatting its code to improve readability and maintainability;

\noindent$\bullet$ \code{FileTabCharacter} and \code{CustomImportOrder} also appear to be frequent violations, which highlights the importance of proper configuration and consistency in code structure; and

\noindent$\bullet$ Violations related to code legibility and ease of reading, such as \code{LineLength} and \code{AvoidStarImport} should not be ignored to maintain a high standard of code quality.

\begin{figure}[!htpb]
	\centering
	\includegraphics[width=0.5\textwidth] {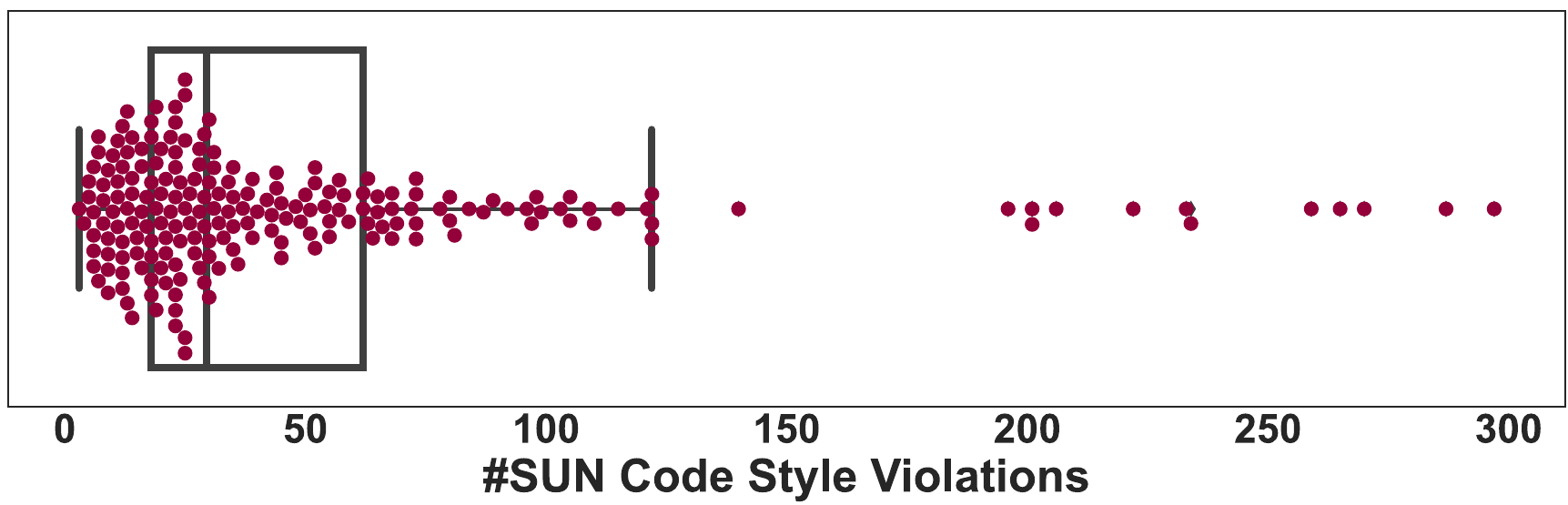}
	\caption{Boxplot for SUN Code Style Violations}
    \label{fig:sun-style}
\end{figure}

\noindent$\triangleright$ \underline{Checkstyle-SUN:} Fig. \ref{fig:sun-style} shows the boxplot. The median value of the data is around 28, with 25\% of the data falling below 15 and 75\% falling below 55. There are several values above the upper quartile, indicating potential outliers or extreme values. The minimum value in the data is 3 and the maximum is 297. The IQR for the dataset is 40, indicating that most of the values in the dataset fall within this range. 

\begin{figure}[!htpb]
	\centering
	\includegraphics[width=0.44\textwidth] {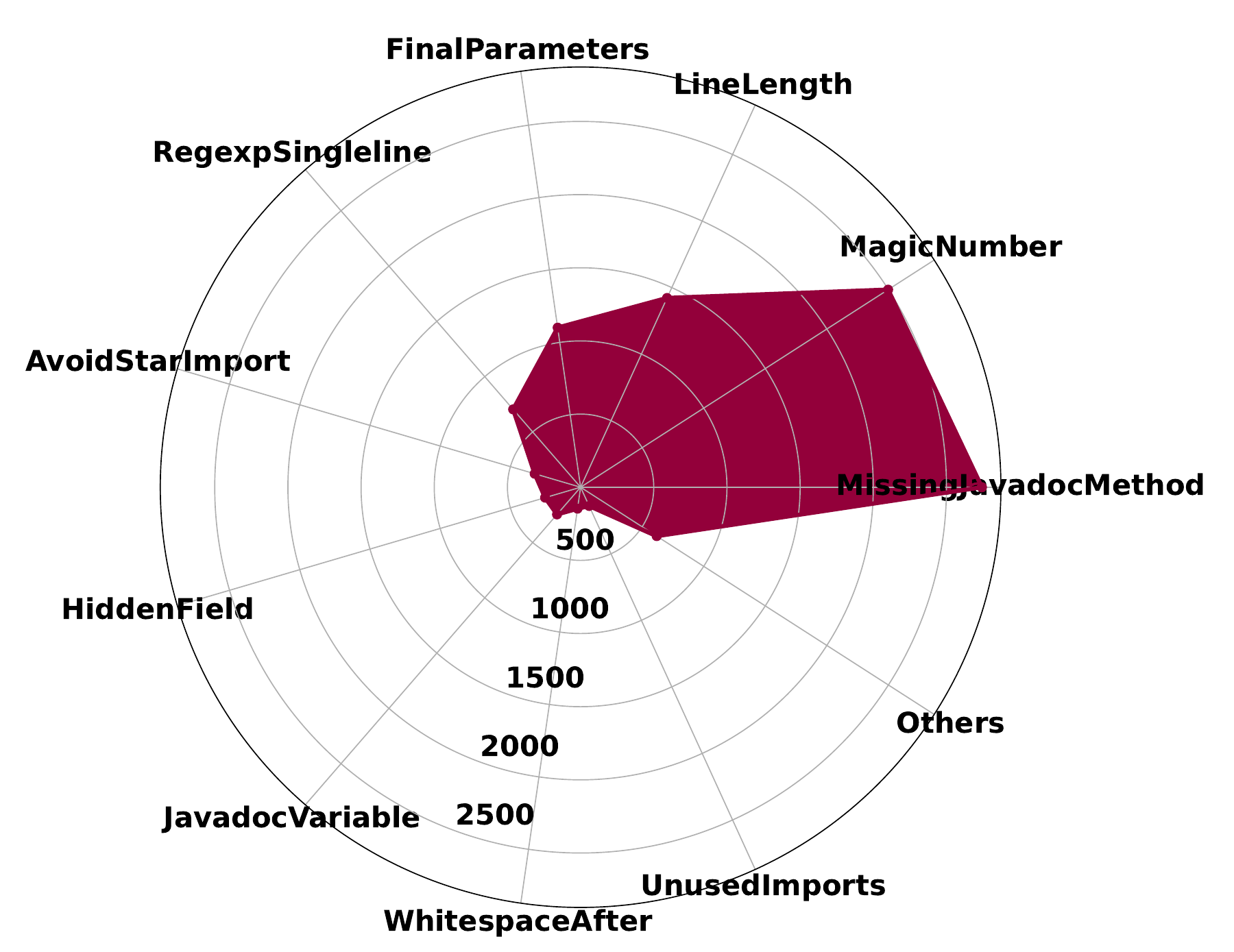}
	\caption{Radar Plot of SUN Code Style Violations}
    \label{fig:sun-style-radar}
\end{figure}

Next, The radar plot in Fig. \ref{fig:sun-style-radar} breakdowns violation issues by types to display the details. As depicted in Fig. \ref{fig:sun-style-radar}, it appears that the two most common types of coding issues are \code{MissingJavadocMethod} and \code{MagicNumber}, with 2742 and 2498 occurrences respectively. The \code{MissingJavadocMethod} issue suggests that more documentation and explanations are required for ChatGPT. Furthermore, magic numbers in the test cases generated by ChatGPT are mainly used in the Assertions. Additionally, the figure shows that \code{FinalParameters}, \code{RegexpSingleline}, and \code{AvoidStarImport} also occur frequently, indicating that attention should be paid to these areas as well. Some of the less frequent issues, such as \code{HiddenField} and \code{UnusedImports}, may be less urgent but still worth addressing to improve overall code quality for ChatGPT. 

In summary, as an AI language model, ChatGPT may not have a specific code style that it adheres to when generating test cases. However, the code style of the test cases can be influenced by the parameters and rules set for the generation process or the input that is given to the model. It also suggests that programmers should pay attention to the code style when using test cases generated by ChatGPT.

\noindent$\bullet$ \textbf{Code Understanding.} The default cyclomatic and cognitive complexity thresholds in PMD are 10 and 15, which means if the cyclomatic and cognitive complexities of a class/method are lower than these values, the system does not report the issue. Thus, we build a series of customized rules to measure complexity. The rule sets can be downloaded from our online repository. Note that, the complexity is measured on a method based.

\noindent$\triangleright$ \underline{Cognitive Complexity:} Based on the technical report from SonarSource \cite{CognitiveComputing}, Cognitive Complexity can be categorized into four categories: low  ($<$5 cognitive complexity), moderate (6-10), high (11-20), and very high complexity (21+). As the results are shown in Table. \ref{tab:cognitivecomplexity}, all methods are with low complexity. 

\begin{table}[!htpb]
\centering
\caption{Cognitive Complexity Results Overview}
\scalebox{0.85}{
\begin{tabular}{ccc}
\hline
\textbf{Cognitive Complexity Level} & \textbf{Num. of Class} & \textbf{Num. of Methods} \\ \hline
Low complexity  ($<$5) & \textcolor[RGB]{0,128,28}{204} & \textcolor[RGB]{0,128,28}{3302}\\
Moderate complexity (6-10) & 0 & 0 \\
High complexity (11-20) & 0 & 0\\ 
Very High complexity (21+) & 0 & 0\\ \hline
\end{tabular}
}
\label{tab:cognitivecomplexity}
\end{table}

\begin{table}[!htpb]
\centering
\caption{Cyclomatic Complexity Results Overview}
\scalebox{0.85}{
\begin{tabular}{ccc}
\hline
\textbf{Cyclomatic Complexity Level} & \textbf{Num. of Class}  & \textbf{Num. of Methods} \\ \hline
Low complexity  (1-4) & \textcolor[RGB]{0,128,28}{204} & \textcolor[RGB]{0,128,28}{3300} \\
Moderate complexity (5-7) & \textcolor[RGB]{0,128,28}{2} & \textcolor[RGB]{0,128,28}{2} \\
High complexity (8-10) & 0 & 0\\ 
Very High complexity (11+) & 0 & 0\\ \hline
\end{tabular}
}
\label{tab:cyclomaticcomplexity}
\end{table}

\noindent$\triangleright$ \underline{Cyclomatic Complexity:} Based on the official documentation from PMD \cite{PMD}, Cognitive Complexity can be categorized into four categories: low  (1-4 cognitive complexity), moderate (5-7), high (8-10), and very high complexity (11+). As the results are shown in Table. \ref{tab:cyclomaticcomplexity}, there are 3300 methods from 204 classes with low complexity and 2 methods from 2 classes with moderate complexity. 

Therefore, based on the aforementioned results, we can conclude that the ChatGPT-generated test cases are overwhelmingly easy to follow and in low complexity.

\begin{tcolorbox}[boxrule=1pt,boxsep=1pt,left=2pt,right=2pt,top=2pt,bottom=2pt,title=Answer to RQ2: Readability]

\noindent$\bullet$ \textbf{Code Style-Google Rule} The median value of approximately 70 (violations). The interquartile range (IQR) falls between around 30 to 175, indicating that most of the data lie within this range. Furthermore, \code{Indentation} is the most common code style violation;

\noindent$\bullet$ \textbf{Code Style-SUN Rule} The median value of the data is around 28 (violations), with 25\% of the data falling below 15 and 75\% falling below 55. The two most common types of coding issues are \code{MissingJavadocMethod} and \code{MagicNumber}, with 2742 and 2498 occurrences respectively; and

\noindent$\bullet$ \textbf{Code Understanding} From the cognitive complexity perspective, all methods are in low complexity. From the cyclomatic complexity perspective, almost all (3300 out of 3302) methods are in low complexity and the other 2 methods are in moderate complexity. Thus, the ChatGPT-generated test cases are overwhelmingly easy to follow and with low complexity.

\end{tcolorbox} 

\subsection{Code Coverage}

\noindent\textbf{RQ3: How does ChatGPT perform with SBST in terms of code coverage?}

\noindent\textbf{Motivation.} While low coverage implies that certain portions of the code have not been checked, high coverage shows that the produced tests have thoroughly evaluated the code. Comparing the code coverage between the test suite generated by ChatGPT and SBST allow us to evaluate and assess the ChatGPT-generated test suite.

\noindent\textbf{Methodology.} The JaCoCo \cite{JaCoCo} measures instruction and branch coverage. The instruction coverage relates to Java bytecode instructions and is thus analogous to statement coverage on source code. We use just instruction coverage (i.e., statement coverage (SC)) to evaluate code coverage as JaCoCo's definition of branch coverage counts only branching of conditional statements, nor edges in the control flow graph. 

\noindent\textbf{Results.} According to the \textit{Long-input setting} in Sec. \ref{subsec:other_setups}, we remove 41 classes and remain 207 Java classes from 75 projects.

\begin{table}[!htpb]
\caption{Statement Code Coverage for Project (I)}
\label{tab:codecoverage}
\centering
\scalebox{0.8}{
\begin{tabular}{l|cccc|c}
\hline
\textbf{Projects} & \textbf{(A)Max} & \textbf{(A)Min} & \textbf{(A)SDEV.} & \textbf{(A)Avg.} & \textbf{(A)ChatGPT} \\ \hline
\textbf{1\_tullibee} & 100\% & 100\% & 0.00 & 100\% & 93\% \\
\textbf{100\_jgaap} & 95.0\% & 95.0\% & 0.00 & 95.0\% & 83\% \\
\textbf{105\_freemind} & 71.5\% & 64.5\% & 1.56 & 69.1\% & 52\% \\
\textbf{107\_weka} & 87.0\% & 79.0\% & 2.95 & 83.0\% &  37\% \\
\textbf{11\_imsmart} & 100\% & 100\% & 0.00 & 100\% & \cellcolor{yellow}{100\%} \\
\textbf{12\_dsachat} & 35.5\% & 35.5\% & 0.00 & 35.5\% & 34\%  \\
\textbf{14\_omjstate} & 67.0\% & 67.0\% & 0.00 & 67.0\% & 55\% \\
\textbf{15\_beanbin} & 80.5\% & 80.5\% & 0.00 & 80.5\% & 46\% \\
\textbf{17\_inspirento} & 94.0\% & 92.5\% & 0.33  & 94.0\% & 87.5\% \\
\textbf{2\_a4j} & 50.5\% & 44.0\% & 1.54 & 48.5\% & 31\% \\
\textbf{21\_geo-google} & 54.0\% & 54.0\% & 0.00 & 54.0\% & \cellcolor{yellow}{67\%} \\
\textbf{24\_saxpath} & 97.0\% & 96.0\% & 0.34 & 96.5\% & 95\% \\
\textbf{26\_jipa} & 88.0\% & 73.5\% & 3.63 & 83.50\% &  \cellcolor{yellow}{97\%} \\
\textbf{29\_apbsmem} & 98.0\% & 98.0\% & 0.00 & 98.0\% & 80\% \\
\textbf{31\_xisemele} & 71.0\% & 71.0\% & 0.00 & 71.0\% & \cellcolor{yellow}{75\%} \\
\textbf{33\_javaviewcontrol} & 82.0\% & 62.5\% & 6.13 &	76.0\% & 46\% \\
\textbf{35\_corina} & 85.0\% & 75.0\% & 3.69 & 78.0\% & 65\% \\
\textbf{36\_schemaspy} & 100.0\% & 100.0\%& 0.00 & 100.0\% & 67\% \\
\textbf{39\_diffi} & 99.0\%	& 93.0\% & 3.02 &95.5\% & 69.5\% \\
\textbf{4\_rif} & 100.0\% & 100.0\%	& 0.00 & 100.00\% & 96\% \\
\textbf{40\_glengineer} & 97.0\% & 86.5\% & 3.37 & 95.0\% & 73\% \\
\textbf{41\_follow} & 92.5\% & 71.0\% & 5.73 & 82.0\% & 38\% \\
\textbf{43\_lilith} & 100.0\% & 100.0\% &	0.00 &	100.0\% & 95\% \\
\textbf{45\_lotus} & 70.5\%	& 70.5\% & 0.00 & 70.5\% &  \cellcolor{yellow}{75\%} \\
\textbf{47\_dvd-homevideo} & 13.3\%	& 13.3\% & 0.00 & 13.3\% &  0.7\% \\
\textbf{51\_jiprof} & 96.5\% & 78.0\% & 3.76 & 93.0\% & 44.5\% \\
\textbf{52\_lagoon} & 19.5\%& 14.0\%& 1.15 & 18.0\% & \cellcolor{yellow}{27\%} \\
\textbf{55\_lavalamp} & 100.0\%	& 100.0\%& 0.00 & 100.0\% & \cellcolor{yellow}{100\%} \\
\textbf{60\_sugar} & 96.0\% & 87.5\%& 2.47 & 90.0\% &  79\% \\
\textbf{61\_noen} & 82.5\% & 81.5\%	& 0.18 & 81.5\% & 60\% \\
\textbf{63\_objectexplorer} & 51.5\% & 51.5\% & 0.00 & 51.5\% & 47\% \\
\textbf{64\_jtailgui} & 76.5\% & 17.0\%	& 16.41 & 70.0\% &  0\% \\
\textbf{68\_biblestudy} & 81.5\% & 81.5\% & 0.00 & 81.5\% & 57\% \\
\textbf{69\_lhamacaw} & 43.5\% & 43.5\% & 0.00 &43.5\% & 6\% \\
\textbf{7\_sfmis} & 100.0\%	& 100.0\%& 0.00 & 100.0\% & 87\% \\
\textbf{72\_battlecry} & 1.0\%	& 1.0\% & 0.00 &1.0\% & \cellcolor{yellow}{57\%} \\
\textbf{73\_fim1} & 24.0\% & 24.0\% & 0.00 &24.0\% &  \cellcolor{yellow}{44.5\%} \\
\textbf{74\_fixsuite} & 67.5\% & 50.0\% & 6.43 & 54.5\% & 40\% \\
\textbf{77\_io-project} & 100.0\% & 100.0\%	& 0.00 & 100.0\% & 71\% \\
\textbf{78\_caloriecount} & 92.7\% & 88.3\%	& 1.24 & 89.7\% & 46.7\% \\
\textbf{79\_twfbplayer} & 97.5\% & 95.5\% & 0.53 & 96.5\% & 69.5\% \\
\textbf{8\_gfarcegestionfa} & 68.0\% & 62.5\% & 1.31 & 65.0\% & 55\% \\
\textbf{80\_wheelwebtool} & 84.3\% & 83.0\% & 0.31 & 83.3\% & 36\% \\
\textbf{82\_ipcalculator} & 91.5\% & 81.0\% & 4.07 & 85.0\% & 73\% \\
\textbf{83\_xbus} & 34.0\% & 19.0\% & 6.75 & 23.00\% &  \cellcolor{yellow}{33\%} \\
\textbf{84\_ifx-framework} &  55.0\% & 55.0\% & 0.00 & 55.0\% & 32\% \\
\textbf{85\_shop} & 71.5\% & 55.8\%	& 4.22 & 63.8\% &  24.8\% \\
\textbf{86\_at-robots2-j} & 86.0\% & 48.0\% & 15.02  &58.0\% & 45\% \\
\textbf{87\_jaw-br} & 32.0\% & 31.0\% & 0.18 &32.0\% & 17.5\% \\
\textbf{88\_jopenchart} & 99.5\% & 72.0\% & 10.87 &	78.5\% & 52\% \\
\textbf{89\_jiggler} & 91.0\% & 81.7\% & 2.10  & 89.7\% & 30.3\% \\
\textbf{90\_dcparseargs} & 100.0\% & 94.0\%	& 1.22 & 99.0\% & 75\% \\
\textbf{91\_classviewer} & 93.0\% & 91.0\% & 0.29 &	92.5\% & 73\% \\
\textbf{92\_jcvi-javacommon} & 100.0\% & 100.0\% & 0.00 &	100.0\% & 74\% \\
\textbf{94\_jclo} & 82.0\% & 68.0\% & 4.31 & 74.0\% & 11\% \\
\textbf{95\_celwars2009} & 47.0\% & 47.0\% & 0.00 & 47.0\% & 46\% \\
\textbf{97\_feudalismgame} & 25.0\% & 19.5\% &	2.06 & 21.1\% &  15\% \\
\textbf{98\_trans-locator} & 50.0\% & 47.0\% & 0.57 & 50.0\% &  15\% \\
\textbf{99\_newzgrabber} &  20.7\% & 17.7\%	& 0.76 & 20.3\%  & 10.7\% \\\hline
\end{tabular}
}
\end{table}

\begin{table}[!htpb]
\caption{Statement Code Coverage for Project (II)}
\label{tab:codecoverage-2}
\centering
\scalebox{0.8}{
\begin{tabular}{l|cccc|c}
\hline
\textbf{Projects} & \textbf{(A)Max} & \textbf{(A)Min} & \textbf{(A)SDEV.} & \textbf{(A)Avg.} & \textbf{(A)ChatGPT} \\ \hline
\textbf{checkstyle} & 87.5\% & 79.3\% & 3.28 &	84.7\% & 65.2\% \\
\textbf{commons-cli} & 98.5\% & 95.0\% & 1.09 & 98.1\% &  69\% \\
\textbf{commons-collections} & 94.3\% & 89.3\%	 & 0.91 &	94.1\% & 68\% \\
\textbf{commons-lang} & 94.0\% & 86.0\%	& 2.36 & 	90.1\%  & 73.1\% \\
\textbf{commons-math} & 72.7\% & 64.1\% & 3.17 & 69.0\% & 45.6\% \\
\textbf{compiler} & 67.7\% & 36.9\% & 9.48 & 53.9\% &  6.29\% \\
\textbf{guava} & 75.0\% & 70.1\% & 1.28 & 72.9\% & 63.1\% \\
\textbf{javaml} & 97.1\% & 87.3\% & 2.46 & 96.4\% & 76.1\% \\
\textbf{javex} & 94.0\% & 67.0\% & 12.59 & 81.2\% & 63\% \\
\textbf{jdom} & 80.7\% & 80.5\% & 0.06 & 80.7\% & 31.3\% \\
\textbf{joda} & 94.9\% & 92.4\% & 0.64 & 93.9\% & 71.6\% \\
\textbf{jsci} & 97.0\% & 86.0\% & 2.62 & 92.4\% & 50\% \\
\textbf{scribe} & 95.3\% & 95.3\% & 0.00 & 95.3\% & 91.2\% \\
\textbf{trove} & 81.0\% & 76.7\% & 1.26 & 79.3\%  & 45.3\% \\
\textbf{twitter4j} & 92.2\% & 89.7\% & 0.67 & 91.3\%  &  70.7\% \\ 
\textbf{xmlenc} & 97.0\% & 94.0\% & 0.61 & 95.1\% & 54\% \\\hline
\textbf{Overall Avg. (Project)} & 77.4\%	& 70.6\% & - & 74.5\% & 55.4\% \\\hline
\end{tabular}
}
\end{table}

\noindent$\triangleright $\textbf{Statement Coverage (SC) Comparison.} As we run 30 times for EvoSuite, we compute the maximum, minimum, average, and average standard deviation. Recall the result in RQ1, for the 3 ChatGPT-generated test cases, which failed to be fixed without the background knowledge, we regard their code coverage as 0 \footnote{Different from 204 test cases in other RQs, we have 207 test cases considered in this RQ.}.

As shown in Table \ref{tab:codecoverage} and \ref{tab:codecoverage-2}, for Evosuite, on average, the maximum SC can reach 77.4\% for all projects; the minimum SC can reach 70.6\% for all projects; and the average SC can reach 74.2\% for all projects. In contrast, for ChatGPT, on average, the average SC can reach 55.4\% for all projects. In general, Evosuite outperforms ChatGPT 19.1\% in regards to SC. Additionally, ChatGPT outperforms Evosuite in 10 out of 75 (13.33\%) projects, which are highlighted in Table. \ref{tab:codecoverage} and \ref{tab:codecoverage-2}. From the class perspective, ChatGPT outperforms EvoSuite in 37 (17.87\%) out of 207 classes. 

Furthermore, by investing 37 cases that ChatGPT outperforms EvoSuite, we find that ChatGPT is highly adept at generating test cases for the following reasons:

\noindent 1. ChatGPT can generate different String objects/integer/double values to use (e.g., comparison) with high diversity compared to Evosuite (Ref: \code{guava::Objects}, \code{math::SimplexTableu});

\noindent 2. ChatGPT can generate 
 an instance of Font for FontChooser, which is not applicable for Evosuite (Ref: \code{71\_film2::FontChooserDialog});

\noindent 3. ChatGPT can generate more reasonable and useable UI operations (i.e., ActionEvents) for testing UIs compared to Evosuite (Ref: \code{72\_bcry::battlecryGUI});

\begin{figure}[!htpb]
	\centering
	\includegraphics[width=0.5\textwidth] {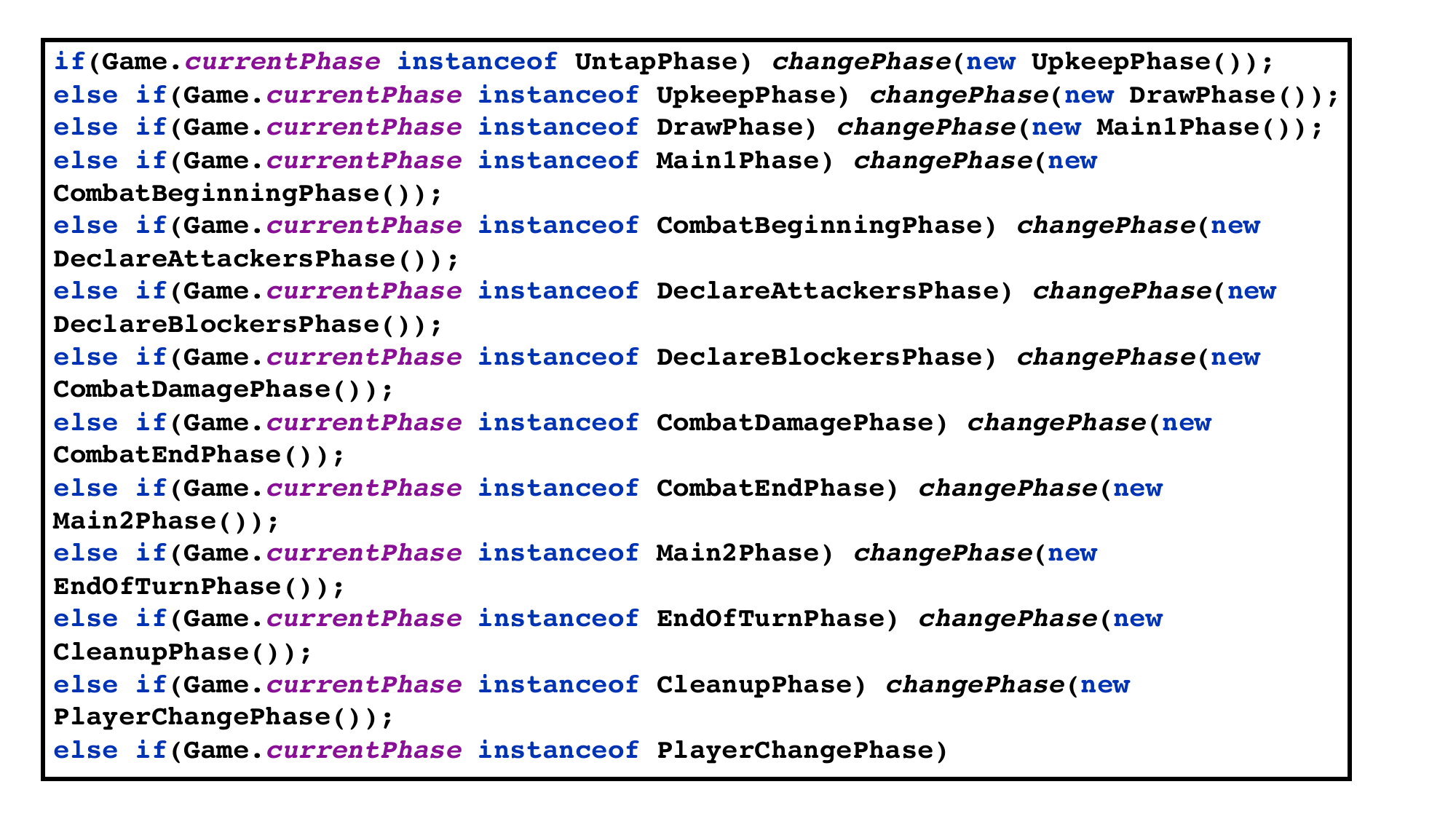}
	\caption{The Code Segment from 45-lotus::Phase}
    \label{fig:lotus-phase}
\end{figure}

\noindent 4. ChatGPT can generate test cases or instances based on the existing information from the classes under tests (Ref: \code{45\_lotus::Phase}). Fig. \ref{fig:lotus-phase} shows a code segment from \code{45-lotus::Phase.java}. This code segment also suggests some instances (e.g, \code{UpkeepPhase()}, \code{DrawPhase()}, \code{Main1Phase()}) are compatible with the type of \newline 
\code{Game.currentPhase}. Such information can be correctly captured by ChatGPT and be used to generate diverse \code{Phase} instances. As a result, it can reach a high coverage than EvoSuite;

\begin{figure}[!htpb]
	\centering
	\includegraphics[width=0.5\textwidth] {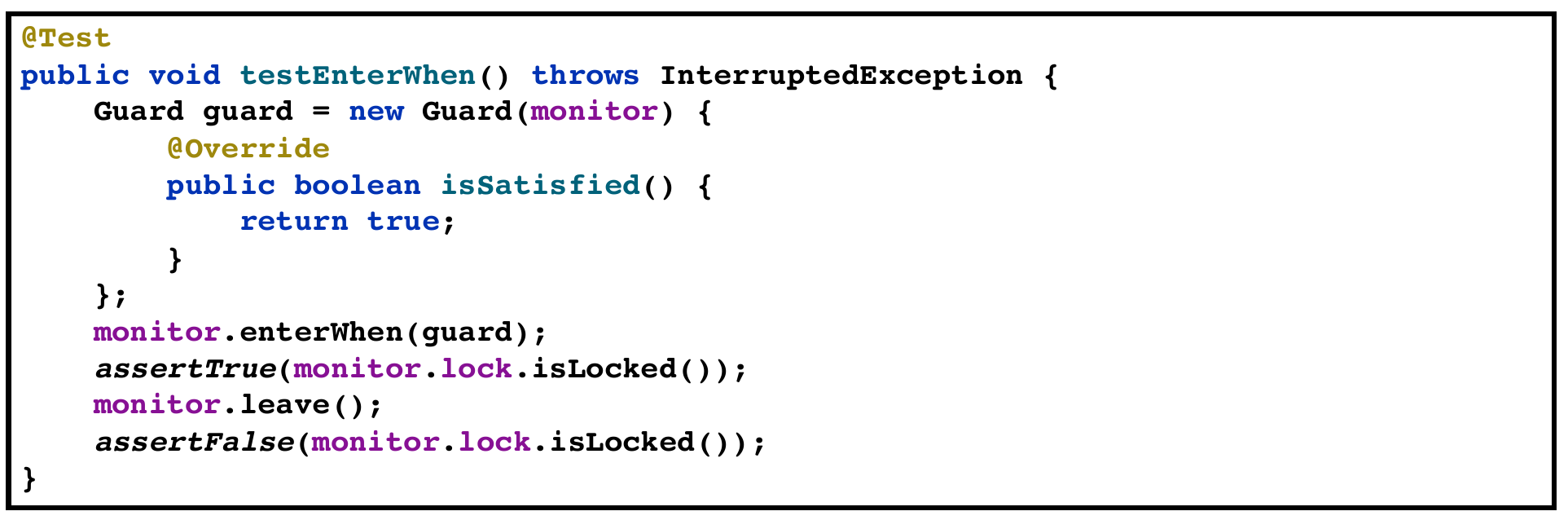}
	\caption{The Code Segment from guava::MonitorTest}
    \label{fig:guava-monitortest}
\end{figure}

\noindent 5. ChatGPT can generate more complex call chains for testing based on the semantics information collected from the classes under test compared to EvoSuite (Ref \code{guava::Monitor}). For example, the code segment in Fig. \ref{fig:guava-monitortest}, ChatGPT can generate a more complex call chain rather than invoking a single method once. More importantly, its call chain is logically correct. That is, the method \code{enter} must be invoked before \code{leave}. This can benefit from that the LLM can precept semantic context from the code or identifiers.

\noindent 6. ChatGPT can generate test data that is suitable for the target regarding the semantic context. For example, the input parameter for invoking the method \code{setCountry} \newline (Ref: \code{21\_geo-google::GeoStatusCode}) can be any String. However, a real country name (e.g., United States) can be more suitable for testing the method \code{setCountry} compared to a random String.

Moreover, as the code complexity increases, so does the search space for identifying appropriate test cases, leading to longer execution times and greater computational expenses for SBST techniques. Consequently, this can pose a significant challenge in uncovering effective test cases that can ensure optimal code coverage and expose any defects. 

Following the previous research works \cite{Zhou:23,Vargha:20,Rojas:2015}, we adopt Vargha-Delaney $\hat{A}_{ab}$ to evaluate
whether a particular approach ($a$) outperforms another ($b$). According to Vargha and Delaney $\hat{A}_{ab}$ 
\cite{Rojas:2015}, negligible, small, medium, and large differences are indicated by A12 over 0.56, 0.64, 0.71, and 0.8, respectively.

\noindent$\triangleright $\textbf{All Classes Comparison.} As shown in Table. \ref{tab:all-class-vd-comparision}, there 193 test cases fall into \textit{large} and 14 test cases fall into \textit{negligible} group. This indicates EvoSuite is overwhelmingly better than ChatGPT in reaching higher code coverage for most cases. The overall Vargha-Delaney measure for all classes is 0.71 (medium).

\begin{table}[!htpb]
\centering
\caption{Vargha-Delaney Measures for Evosuite vs. ChatGPT}
\begin{tabular}{ccccc}
\hline
 & \textbf{Large} & \textbf{Medium} & \textbf{Small} & \textbf{Negligible} \\ \hline
Num. of Classes & 193 & 0 & 0 & 14 \\ \hline
Overall V.D. &  \multicolumn{4}{c}{0.71 (Medium)} \\ \hline
\end{tabular}
\label{tab:all-class-vd-comparision}
\end{table}

\noindent$\triangleright $\textbf{Small/Big Classes Comparison.} Here, small classes are defined as classes with less than 50 branches. Classes with more than 50 branches are considered as big classes. 

\begin{table}[!htpb]
\centering
\caption{Vargha-Delaney Measures for Big Classes}
\begin{tabular}{ccccc}
\hline
 & \textbf{Large} & \textbf{Medium} & \textbf{Small} & \textbf{Negligible} \\ \hline
Num. of Big Classes & 121 & 0 & 0 & 5 \\ \hline
Overall V.D. &  \multicolumn{4}{c}{0.764 (Large)} \\ \hline
\end{tabular}
\label{tab:all-big-vd-comparision}
\end{table}

\begin{table}[!htpb]
\centering
\caption{Vargha-Delaney Measures for Small Classes}
\begin{tabular}{ccccc}
\hline
 & \textbf{Large} & \textbf{Medium} & \textbf{Small} & \textbf{Negligible} \\ \hline
Num. of Small Classes & 70 & 0 & 0 & 11 \\ \hline
Overall V.D. &  \multicolumn{4}{c}{0.63 (Small)} \\ \hline
\end{tabular}
\label{tab:all-small-vd-comparision}
\end{table}

\noindent$\triangleright $\textbf{Big Classes Comparison.} Table. \ref{tab:all-big-vd-comparision} shows the comparison for big classes. There 121 test cases fall into \textit{large} and 5 test cases fall into \textit{negligible} group. This indicates EvoSuite is overwhelmingly better than ChatGPT in reaching higher code coverage for big class cases. The overall Vargha-Delaney measure for all classes is 0.764 (large).

\noindent$\triangleright $\textbf{Small Classes Comparison.} Table. \ref{tab:all-small-vd-comparision} shows the comparison for small classes. There 70 test cases fall into \textit{large} and 11 test cases fall into \textit{negligible} group. This indicates EvoSuite is overwhelmingly better than ChatGPT in reaching higher code coverage for big class cases. The overall Vargha-Delaney measure for all classes is 0.63 (small).

Unfortunately, we fail to see ChatGPT outperforms EvoSuite for even big classes. It indicates no matter the big or small classes, developers are suggested to turn to EvoSuite in order to obtain a higher code coverage. The potential causes may be diverse and varied. Some possible reasons can be: (1) \textbf{incomplete specifications:} ChatGPT is only given the classes under test instead of the entire project. Thus, without the information from the entire project, it can be hard for ChatGPT to generate more valuable test cases; (2) \textbf{lack of feedback mechanisms:} Unlike Evosuit, which can learn from feedback (i.e., cover data), ChatGPT relies solely on the training data. It makes it challenging for ChatGPT to comprehend the feedback from test results through an iterative process leading to low test coverage.

However, the results also suggest two insights:

\noindent $\star $\textbf{Insight 1:} As an AI-powered assistant, ChatGPT has a strong capability in understanding precept semantics and context from the code under test. This means that ChatGPT can assist in generating test data effectively. By embedding an AI model or an NLP (Natural Language Processing) module within an SBST (Search-Based Software Testing) tool, ChatGPT can greatly improve the performance of the SBST tool. This is because the tool will be able to comprehend and interpret complicated code structures and generate test cases based on them with higher accuracy and efficiency. As a result, developers can benefit from faster, more efficient testing, and a more reliable software product; and

\noindent $\star $\textbf{Insight 2:} Even though it cannot compare with EvoSuite, ChatGPT can still reach a relatively high code coverage (55.4\%). Thus, ChatGPT can still serve as an entry-level tool for testing newcomers or as a backup option.

\begin{tcolorbox}[boxrule=1pt,boxsep=1pt,left=2pt,right=2pt,top=2pt,bottom=2pt,title=Answer to RQ3: Code Coverage]
\noindent$\bullet$ For Evosuite, on average, the maximum SC can reach 77.4\% for all projects; the minimum SC can reach 70.6\%; and the average SC can reach 74.2\%. In contrast, for ChatGPT, on average, the average SC can reach 55.4\%;

\noindent$\bullet$ After examining 37 cases in which ChatGPT outperformed EvoSuite (in code coverage), our analysis suggests six potential scenarios where ChatGPT may be better suited. These findings contribute to a growing body of research exploring the efficacy of automated testing tools;

\noindent$\bullet$ The experimental results indicate EvoSuite is overwhelmingly better than ChatGPT in reaching higher code coverage for both big class cases and small class cases; and 

\noindent$\bullet$ Two potential reasons for low code coverage can be: incomplete specifications; and lack of feedback mechanisms.
\end{tcolorbox} 

\subsection{Bug Detection}
\noindent\textbf{RQ4: How effective are ChatGPT and SBST in generating test suites that detect bugs?}

\noindent\textbf{Motivation.} The main use of generated test suites is finding buddy code in a program. Therefore, in this RQ, we evaluate the effectiveness of generated test suite in detecting bugs.

\noindent\textbf{Methodology.} To evaluate the effectiveness of generated test suite in terms of detecting bugs, we first generate unit test suites for the target classes and examine whether the test suite can successfully capture the bug in the Defects4J benchmark. Note that, in this RQ, for fairness, we only run EvoSuite once to generate test cases. 

\begin{figure}[!htpb]
	\centering
	\includegraphics[width=0.5\textwidth] {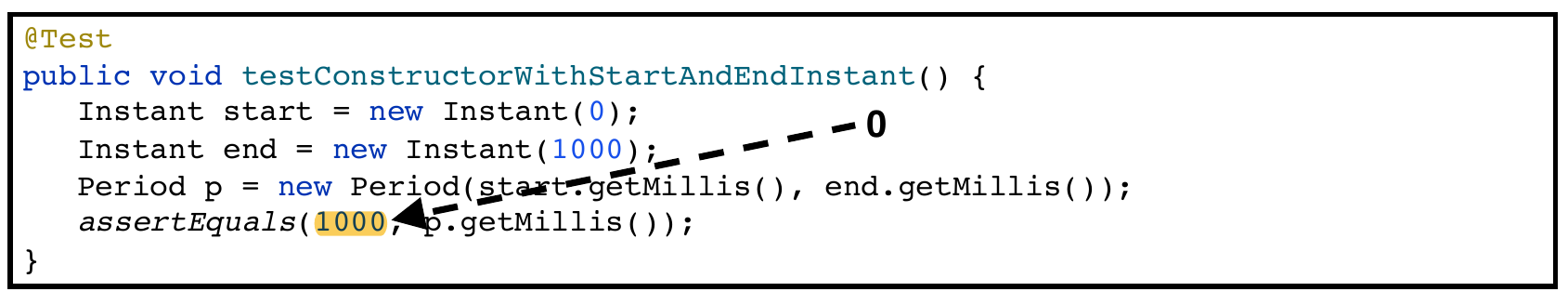}
	\caption{The Test Cases for Time Project}
    \label{fig:time-buggy}
\end{figure}

\noindent\textbf{Results.}
Some bugs in the Defects4J are logical bugs, which are triggered with \code{Assertions}. Unfortunately, we find that sometimes the \code{Assertions} generated by ChatGPT are not reliable. For example, Fig. \ref{fig:time-buggy} illustrates a test case for \code{Period} in Time project. The Assertion statement \code{assertEquals(1000, p.getMillis();} is incorrect. However, the code segment under test is not buggy and the expected value should be 0 instead of 1000. ChatGPT makes an incorrect Assertion for this case. It means we cannot fully rely on the \code{Assertions} in ChatGPT-generated test cases to determine whether the bugs are successfully triggered. However, manually checking the Assertions in ChatGPT-generated test cases can be effort-consuming and error-prone \cite{Kavir:2011,Gunel:21,Valerio:21}. Therefore, in this RQ, we focus on bugs that associate with Java Exceptions, such as \code{NullPointerException}, \code{UnsupportedOperationException}. 

\begin{table}[!htpb]
\caption{Bug Detection Comparison for ChatGPT and Evosuite}
\centering
\label{tab:bug-detection}
\scalebox{0.72}{
\begin{tabular}{cccccc}
\hline
 & \multicolumn{1}{c|}{} & \multicolumn{2}{c|}{\textbf{ChatGPT}} & \multicolumn{2}{c}{\textbf{Evosuite}} \\ \hline
\textbf{Project} & \multicolumn{1}{c|}{\textbf{\# All/ \# Exce. Bugs}} & \textbf{Detected} & \multicolumn{1}{c|}{\textbf{ Coverage}} & \textbf{Detected} & \textbf{ Coverage} \\ \hline
Chart & 26 / 8 & \textcolor[RGB]{255,0,94}{4 (50\%)} & 62\% & 3 (38\%) & \textcolor[RGB]{255,0,94}{85\%} \\ 
Cli & 39 / 8 & 1 (13\%) & 70\% & \textcolor[RGB]{255,0,94}{2  (25\%)} & \textcolor[RGB]{255,0,94}{88\%}  \\ 
Closure & 174 / 9 & \textcolor[RGB]{255,0,94}{1 (11\%)} & \textcolor[RGB]{255,0,94}{14\%} & 0 (0\%) & 4\% \\ 
Codec & 18 / 7 & 0 (0\%) & 60\% & \textcolor[RGB]{255,0,94}{2 (29\%)} & \textcolor[RGB]{255,0,94}{94\%} \\ 
Collections & 4 / 2 & 0 (0\%) & \textcolor[RGB]{255,0,94}{87\%} & 0 (0\%) & 67\% \\ 
Compress & 47 / 19 & \textcolor[RGB]{255,0,94}{6 (32\%)} & 42\% & 3 (16\%) & \textcolor[RGB]{255,0,94}{57\%} \\ 
Csv & 16 / 7 & 2 (29\%) & 80\% & \textcolor[RGB]{255,0,94}{5 (71\%)} & \textcolor[RGB]{255,0,94}{90\%} \\ 
Gson & 18 /12 & 2 (17\%) & \textcolor[RGB]{255,0,94}{59\%} &  \textcolor[RGB]{255,0,94}{6 (50\%)} & 55\% \\ 
JacksonCore & 26 / 8 & 2 (25\%) & 38\% & 2 (25\%) & \textcolor[RGB]{255,0,94}{64\%} \\ 
JacksonDatabind & 112 / 53 & 9 (17\%) & 30\% & \textcolor[RGB]{255,0,94}{4 (8\%)} &  \textcolor[RGB]{255,0,94}{56\%} \\ 
JacksonXml & 6 / 1 &  0 (0\%) & 29\% & 0 (0\%) & 49\% \\ 
Jsoup & 93 / 22 & 4 (18\%) & 63\% & \textcolor[RGB]{255,0,94}{10 (45\%)} & \textcolor[RGB]{255,0,94}{86\%} \\ 
JxPath & 22 / 1 & 1 (100\%) & 40\% & 1 (100\%) & \textcolor[RGB]{255,0,94}{88\%} \\ 
Lang & 64 / 20 & \textcolor[RGB]{255,0,94}{6 (30\%)} & \textcolor[RGB]{255,0,94}{68\%} & 3 (15\%) & 55\% \\ 
Math & 106 / 28 & 5 (18\%) & 64\% & \textcolor[RGB]{255,0,94}{12 (43\%)}  & \textcolor[RGB]{255,0,94}{84\%} \\ 
Time & 26 / 7  & 1 (14\%) & 56\% & \textcolor[RGB]{255,0,94}{2 (29\%)} & \textcolor[RGB]{255,0,94}{88\%}  \\ \hline
Total & 796 / 212 &  44 (21\%) & 50\% & 55 (26\%)  & 67\% \\ 
\hline
\end{tabular}
}
\end{table}

Table. \ref{tab:bug-detection} shows the experimental results. In the table, for each project, the higher values (e.g., higher code coverage) are highlighted in comparison between the two approaches. Furthermore, out of 212 bugs, 44 were successfully detected by test cases generated by ChatGPT, with an average statement code coverage of 50\%. In contrast, test cases generated by EvoSuite successfully detected 55 bugs, with an average statement code coverage of 67\%. From the comparison, we can also see that in some cases, EvoSuite detected more bugs than ChatGPT, while in other cases, ChatGPT detected more bugs than EvoSuite. For example, in the Chart project, EvoSuite had a higher coverage rate for bug detection than ChatGPT, but ChatGPT detected more bugs than EvoSuite in some cases. It is worth noting that the coverage rates for both tools varied greatly across different projects, indicating that the effectiveness of each tool may depend on the specific characteristics of the project being tested. It is interesting to note that ChatGPT was able to detect bugs in some cases where EvoSuite was not, indicating that the two tools may complement each other and could be used together to improve bug detection.

By comparing the test cases generated by ChatGPT and EvoSuite, we find several possible reasons that LLM (e.g., ChatGPT) may not outperform Evosuite:

\noindent$\bullet$ As the input for the ChatGPT can only the class under test instead of the entire project (e.g. jar file), it can be hard for ChatGPT generate complex instances, which can make the test cases to generate corner cases to explore bugs;

\noindent$\bullet$ As a large language model, ChatGPT generates/predicts content takes a prompt or starting text as input, and uses its learned understanding of language to predict what words or phrases should come next. This prediction is based on the probability that a certain sequence of words would appear in the dataset. It is highly possible that a commonly used case (i.e., test case/data in our context) holds a higher probability compared to an edge case; and

\noindent$\bullet$ By adopting the genetic algorithm to explore potential test suites capable of achieving higher code coverage, Evosuite may theoretically possess a greater probability of discovering bugs. Notably, such a feedback mechanism is presently absent in LLMs, such as ChatGPT, underscoring the potential benefits of combining SBST techniques with LLMs for program testing and bug detection.

It is also worth mentioning that the results presented do not reflect the capability of ChatGPT in finding or locating bugs. It only implicates the bug detection capability of ChatGPT-generated test cases.

\begin{tcolorbox}[boxrule=1pt,boxsep=1pt,left=2pt,right=2pt,top=2pt,bottom=2pt,title=Answer to RQ4: Defects and Bug Detection]

\noindent$\bullet$ The test cases generated by ChatGPT can be misleading in finding logical-related bugs, as the \code{Assertions} generated can be incorrect and unreliable;

\noindent$\bullet$ Out of 212 bugs, 44 were successfully detected by test cases generated by ChatGPT, with an average statement code coverage of 50\%. In contrast, test cases generated by EvoSuite successfully detected 55 bugs, with an average statement code coverage of 67\%;

\noindent$\bullet$ Evosuite integrates a genetic algorithm to find test cases that can provide better code coverage and increase the chances of finding bugs. LLM tools like ChatGPT do not have this feedback mechanism. Thus, combining the SBST technique and LLM can improve software testing accuracy and bug detection.

\end{tcolorbox}

\section{Limitations, and Threats to Validity}
\subsection{Limitations}
The results and experiments of this study is limited in two parts: (1) Given the need of manually query ChatGPT, our study is limited to only the queries made for the study. As ChatGPT is a closed-source and we cannot map our results to the details or characteristics of ChatGPT’s internal model. We also do not know ChatGPT’s exact training data, which means we cannot determine if the exact response to our queries are members of the training data; and (2) As ChatGPT is continuously updating and training, the responses of ChatGPT can only reflect the performance of ChatGPT at the time we conduct our work (i.e., ChatGPT Jan 30 (2023) Version).

\subsection{Threats to Validity}
To reduce bias by manually selecting subject programs for testing, we reuse the benchmarks (i.e., Defects4J, DyanMOSA Dataset), which have been used and studied in the existing researches. Furthermore, we also reuse the metrics presented in existing research works to calculate the code coverage, code readability and so forth. Another threat to internal validity comes from the randomness of the genetic algorithms. To reduce the risk, we repeat EvoSuite for 30 times for every class. As for external validity, due to size of the benchmarks, we do not attempt to generalize our results and conclusions.
\section{Related Work}

\noindent\textbf{Language Models.} Language models are used in NLP for many tasks, such as, machine translation, question answering, summarization, text generation and so on~\cite{Carlini:21,Raffel:22,Lan:19,zhang:20,Pilault:20,Cai:21,Khashabi:20,Cho:14,Chen:21,Bui:21}. To better understand language, models with massive parameters are trained on an extremely large corpus (i.e., LLM). Transformer~\cite{Vaswani:17} is constructed on stacked encoders and decoders. It leverages self-attention mechanism to weigh the importance of words in the input text, capturing long-range dependencies and relationships between words in the input. It is the base for many LLMs. ELMo~\cite{Peters:18} utilizes multi-layer bidirectional LSTM and provides high-quality word representations. GPT~\cite{radford:18} and BERT~\cite{Devlin:18} are built on the decoders (unidirectional) and encoders (bidirectional) of Transformer, respectively, using pre-training and fine-tuning techniques. GPT-2~\cite{radford:19} and GPT-3~\cite{Brown:20} are the descendants of GPT. GPT-2 has a larger model size than GPT, and GPT-3 is larger than GPT-2. Moreover, with larger corpus, GPT-2 and GPT-3 introduce zero-shot and few-shot learning to make models adapt to Multitask. Codex~\cite{Chen:21} is obtained by training GPT-3 using Github code data. It is the model that powers GitHub Copilot~\cite{Copilot}, a tool generating computer code automatically. InstructGPT~\cite{Ouyang:22} utilizes additional supervised learning and reinforcement learning from human feedback to fine-tune GPT-3, aligning LLM with users. ChatGPT~\cite{ChatGPT} uses the same methods as InstructGPT and has the ability to answer follow-up questions.

\noindent\textbf{Search-based Software Testing.} SBST approaches test case generation as an optimization problem. The first SBST method to produce test data for functions with float-type inputs was put out by Miller et al. ~\cite{Miller:76}. Many software testing methods \cite{Li:07, Silva:17, Walcott:06} have made extensive use of SBST approaches. Most studies concentrate on (1) Search algorithms: Tonella \cite{Tonella:04} suggested iterating to generate one test case for each branch. A test suite for all branches was suggested by Fraser et al. \cite{Fraser:13}. Many-objective optimization techniques were presented by Panichella et al. \cite{Panichella:15, Panichella:18}. To lower the expenses of computing, Grano et al. \cite{Grano:19} developed a variation of DynaMOSA; (2) Enhancing fitness gradients: Arcuri et al. introduced testability transformations into API tests \cite{Arcuri:21} For programs with complicated inputs. Lin et al. \cite{Lin:20} suggested an approach to deal with the inter-procedural flag issue. A test seed synthesis method was suggested by Lin et al. to produce complicated test inputs \cite{Lin:21}. Braione et al. \cite{Pietro:17} coupled symbolic execution and SBST; (3) Design of the fitness function: Xu et al. \cite{Xu:17} suggested an adaptive fitness function for enhancing SBST;  Rojas et al. \cite{Miguel:15} suggested combining multiple coverage criteria for fulfilling more requirements from developers. Gregory Gay experimented with various criterion combinations \cite{Gay:17} to compare the usefulness of multi-criteria suites for spotting practical flaws. Zhou et al. \cite{Zhou:23} proposed a method to select coverage goals from multiple criteria instead of combining all goals; (4) Readability of created tests: Daka et al. \cite{Daka:17} suggested naming tests by stating covered goals. Deep learning techniques were presented by Roy et al. \cite{Roy:20}; (5) Applying SBST to more software fields such as Machine Learning libraries \cite{Wang:21}, Android applications \cite{Dong:2020}, Web APIs \cite{Martin:2021}, and Deep Neural Networks \cite{Haq:2021}.

\section{Conclusion}
In this article, we present a systematic assessment of unit test suites generated by two state-of-the-art techniques: ChatGPT and SBST. We comprehensively evaluate test suites generated by ChatGPT from multiple critical perspectives, including correctness, readability, code coverage, and bug detection capability. Our experimental results demonstrate that (1) 69.6\% of the ChatGPT-generated test cases can be successfully compiled and executed; (2) We also observed that the most common violations in the generated code style were \code{Indentation} (for Google Style) and \code{MissingJavadocMethod} (for SUN Style), while the majority of the test cases exhibited low complexity; (3) Moreover, our evaluation revealed that EvoSuite outperforms ChatGPT in terms of code coverage by 19\%; and (4) EvoSuite outperforms ChatGPT in terms of code coverage by 5\%.
\section{Data Availability}\label{sec:data}
The experimental results and raw data are available at: \url{https://sites.google.com/view/chatgpt-sbst}

\bibliographystyle{IEEEtran}

\begin{thebibliography}{10}
\providecommand{\url}[1]{#1}
\csname url@samestyle\endcsname
\providecommand{\newblock}{\relax}
\providecommand{\bibinfo}[2]{#2}
\providecommand{\BIBentrySTDinterwordspacing}{\spaceskip=0pt\relax}
\providecommand{\BIBentryALTinterwordstretchfactor}{4}
\providecommand{\BIBentryALTinterwordspacing}{\spaceskip=\fontdimen2\font plus
\BIBentryALTinterwordstretchfactor\fontdimen3\font minus
  \fontdimen4\font\relax}
\providecommand{\BIBforeignlanguage}[2]{{%
\expandafter\ifx\csname l@#1\endcsname\relax
\typeout{** WARNING: IEEEtran.bst: No hyphenation pattern has been}%
\typeout{** loaded for the language `#1'. Using the pattern for}%
\typeout{** the default language instead.}%
\else
\language=\csname l@#1\endcsname
\fi
#2}}
\providecommand{\BIBdecl}{\relax}
\BIBdecl

\bibitem{Zhu:1997}
H.~Zhu, P.~A.~V. Hall, and J.~H.~R. May, ``Software unit test coverage and
  adequacy,'' \emph{ACM Comput. Surv.}, vol.~29, no.~4, p. 366–427, 1997.

\bibitem{Mark:12}
M.~Harman, S.~A. Mansouri, and Y.~Zhang, ``Search-based software engineering:
  Trends, techniques and applications,'' \emph{ACM Computing Surveys (CSUR)},
  vol.~45, no.~1, pp. 1--61, 2012.

\bibitem{Fraser:13}
G.~Fraser and A.~Arcuri, ``Whole test suite generation,'' \emph{IEEE
  Transactions on Software Engineering}, vol.~39, no.~2, pp. 276--291, 2013.

\bibitem{Zhou:23}
Z.~Zhou, Y.~Zhou, C.~Fang, Z.~Chen, and Y.~Tang, ``Selectively combining
  multiple coverage goals in search-based unit test generation,'' in \emph{37th
  IEEE/ACM International Conference on Automated Software Engineering}, 2022,
  pp. 1--12.

\bibitem{Carlini:21}
N.~Carlini, F.~Tramer, E.~Wallace, M.~Jagielski, A.~Herbert-Voss, K.~Lee,
  A.~Roberts, T.~B. Brown, D.~Song, U.~Erlingsson \emph{et~al.}, ``Extracting
  training data from large language models.'' in \emph{USENIX Security
  Symposium}, vol.~6, 2021.

\bibitem{Brants:07}
T.~Brants, A.~C. Popat, P.~Xu, F.~J. Och, and J.~Dean, ``Large language models
  in machine translation,'' 2007.

\bibitem{Raffel:22}
C.~Raffel, N.~Shazeer, A.~Roberts, K.~Lee, S.~Narang, M.~Matena, Y.~Zhou,
  W.~Li, and P.~J. Liu, ``Exploring the limits of transfer learning with a
  unified text-to-text transformer,'' \emph{J. Mach. Learn. Res.}, vol.~21,
  no.~1, 2022.

\bibitem{Svyatkovskiy:20}
A.~Svyatkovskiy, S.~K. Deng, S.~Fu, and N.~Sundaresan, ``Intellicode compose:
  Code generation using transformer,'' in \emph{Proc. of ESEC/FSE}, 2020, p.
  1433–1443.

\bibitem{Alon:20}
U.~Alon, R.~Sadaka, O.~Levy, and E.~Yahav, ``Structural language models for
  any-code generation,'' 2019.

\bibitem{Poesia:22}
G.~Poesia, A.~Polozov, V.~Le, A.~Tiwari, G.~Soares, C.~Meek, and S.~Gulwani,
  ``Synchromesh: Reliable code generation from pre-trained language models,''
  in \emph{Proc. of ICLR}, 2022.

\bibitem{McBurney:16}
P.~W. McBurney and C.~McMillan, ``Automatic source code summarization of
  context for java methods,'' \emph{IEEE Transactions on Software Engineering},
  vol.~42, no.~2, pp. 103--119, 2016.

\bibitem{Haiduc:10}
S.~Haiduc, J.~Aponte, and A.~Marcus, ``Supporting program comprehension with
  source code summarization,'' in \emph{Proc. of ICSE}, 2010, p. 223–226.

\bibitem{Zhang:20ICSE}
J.~Zhang, X.~Wang, H.~Zhang, H.~Sun, and X.~Liu, ``Retrieval-based neural
  source code summarization,'' in \emph{Proc. of ICSE}, 2020, p. 1385–1397.

\bibitem{McBurney:14}
P.~W. McBurney and C.~McMillan, ``Automatic documentation generation via source
  code summarization of method context,'' in \emph{Proc. of ICPC}, 2014, p.
  279–290.

\bibitem{Hu:18}
X.~Hu, G.~Li, X.~Xia, D.~Lo, and Z.~Jin, ``Deep code comment generation,'' in
  \emph{Proc. of ICPC}, 2018, p. 200–210.

\bibitem{Brown:20}
T.~Brown, B.~Mann, N.~Ryder, M.~Subbiah, J.~D. Kaplan, P.~Dhariwal,
  A.~Neelakantan, P.~Shyam, G.~Sastry, A.~Askell, S.~Agarwal, A.~Herbert-Voss,
  G.~Krueger, T.~Henighan, R.~Child, A.~Ramesh, D.~Ziegler, J.~Wu, C.~Winter,
  C.~Hesse, M.~Chen, E.~Sigler, M.~Litwin, S.~Gray, B.~Chess, J.~Clark,
  C.~Berner, S.~McCandlish, A.~Radford, I.~Sutskever, and D.~Amodei, ``Language
  models are few-shot learners,'' in \emph{Advances in Neural Information
  Processing Systems}, vol.~33, 2020, pp. 1877--1901.

\bibitem{ChatGPT}
OpenAI, ``Chatgpt: Optimizing language models for dialogue,'' 2023,
  https://openai.com/blog/chatgpt/.

\bibitem{Tonella:04}
P.~Tonella, ``Evolutionary testing of classes,'' in \emph{Proc. of ISSTA},
  2004, p. 119–128.

\bibitem{Panichella:15}
A.~Panichella, F.~M. Kifetew, and P.~Tonella, ``Reformulating branch coverage
  as a many-objective optimization problem,'' in \emph{Proc. of ICST}, 2015,
  pp. 1--10.

\bibitem{Panichella:18}
------, ``Automated test case generation as a many-objective optimisation
  problem with dynamic selection of the targets,'' \emph{IEEE Transactions on
  Software Engineering}, vol.~44, no.~2, pp. 122--158, 2018.

\bibitem{Fraser:11}
G.~Fraser and A.~Arcuri, ``Evosuite: Automatic test suite generation for
  object-oriented software,'' in \emph{Proc. of ESEC/FSE}, 2011, p. 416–419.

\bibitem{Vaswani:17}
A.~Vaswani, N.~Shazeer, N.~Parmar, J.~Uszkoreit, L.~Jones, A.~N. Gomez,
  {\L}.~Kaiser, and I.~Polosukhin, ``Attention is all you need,''
  \emph{Advances in neural information processing systems}, vol.~30, 2017.

\bibitem{Devlin:18}
J.~Devlin, M.-W. Chang, K.~Lee, and K.~Toutanova, ``Bert: Pre-training of deep
  bidirectional transformers for language understanding,'' \emph{arXiv preprint
  arXiv:1810.04805}, 2018.

\bibitem{Raffel:20}
C.~Raffel, N.~Shazeer, A.~Roberts, K.~Lee, S.~Narang, M.~Matena, Y.~Zhou,
  W.~Li, and P.~J. Liu, ``Exploring the limits of transfer learning with a
  unified text-to-text transformer,'' \emph{The Journal of Machine Learning
  Research}, vol.~21, no.~1, pp. 5485--5551, 2020.

\bibitem{Ouyang:22}
L.~Ouyang, J.~Wu, X.~Jiang, D.~Almeida, C.~L. Wainwright, P.~Mishkin, C.~Zhang,
  S.~Agarwal, K.~Slama, A.~Ray \emph{et~al.}, ``Training language models to
  follow instructions with human feedback,'' \emph{arXiv preprint
  arXiv:2203.02155}, 2022.

\bibitem{Artetxe:22}
M.~Artetxe, J.~Du, N.~Goyal, L.~Zettlemoyer, and V.~Stoyanov, ``On the role of
  bidirectionality in language model pre-training,'' \emph{arXiv preprint
  arXiv:2205.11726}, 2022.

\bibitem{radford:19}
A.~Radford, J.~Wu, R.~Child, D.~Luan, D.~Amodei, I.~Sutskever \emph{et~al.},
  ``Language models are unsupervised multitask learners,'' \emph{OpenAI blog},
  vol.~1, no.~8, p.~9, 2019.

\bibitem{radford:18}
A.~Radford, K.~Narasimhan, T.~Salimans, I.~Sutskever \emph{et~al.}, ``Improving
  language understanding by generative pre-training,'' 2018.

\bibitem{Lin:21}
Y.~Lin, Y.~S. Ong, J.~Sun, G.~Fraser, and J.~S. Dong, ``Graph-based seed object
  synthesis for search-based unit testing,'' in \emph{Proc. of ESEC/FSE}, 2021,
  p. 1068–1080.

\bibitem{Defects4J}
Defects4J, ``Defects4j: A database of real faults and an experimental
  infrastructure to enable controlled experiments in software engineering
  research,'' 2023, https://github.com/rjust/defects4j.

\bibitem{Evosuite}
Evosuite, ``Evosuite: Automatic test suite generation for java,'' 2023,
  https://www.evosuite.org/.

\bibitem{SpotBugs}
SpotBugs, ``Spotbugs,'' 2023, https://spotbugs.github.io/index.html.

\bibitem{Findbugs}
B.~Pugh and D.~Hovemeyer, ``Findbugs,'' 2023,
  https://findbugs.sourceforge.net/.

\bibitem{Ayewah:08}
N.~Ayewah, W.~Pugh, D.~Hovemeyer, J.~D. Morgenthaler, and J.~Penix, ``Using
  static analysis to find bugs,'' \emph{IEEE Software}, vol.~25, no.~5, pp.
  22--29, 2008.

\bibitem{IntelliJ}
JetBrain, ``Intellij idea – the leading java and kotlin ide,'' 2023,
  https://www.jetbrains.com/idea/.

\bibitem{SpotBugsDesc}
Spotbugs, ``Spotbug bug descriptions,'' 2023,
  https://spotbugs.readthedocs.io/en/stable/bugDescriptions.html.

\bibitem{CheckStyle}
CheckStyle, ``Checkstyle,'' 2023, https://checkstyle.sourceforge.io/.

\bibitem{SUNJavaStyle}
Oracle, ``Code conventions for the java programming language,'' 1999,
  https://www.oracle.com/java/technologies/javase/codeconventions-contents.html.

\bibitem{GoogleJavaStyle}
Google, ``Google java style guide,'' 2023,
  https://google.github.io/styleguide/javaguide.html.

\bibitem{Dantas:21}
C.~E.~C. Dantas and M.~A. Maia, ``Readability and understandability scores for
  snippet assessment: an exploratory study,'' \emph{arXiv preprint
  arXiv:2108.09181}, 2021.

\bibitem{PMD}
P.~S.~C. Analyzer, ``Pmd,'' 2023, https://pmd.github.io/.

\bibitem{CognitiveComputing}
sonarsource, ``Cognitive computing: A new way of measuring understandability,''
  2021, https://www.sonarsource.com/docs/CognitiveComplexity.pdf.

\bibitem{JaCoCo}
M.~G. .~C. KG, ``Jacoco java code coverage library,'' 2023,
  https://www.jacoco.org/jacoco/.

\bibitem{Vargha:20}
A.~Vargha and H.~D. Delaney, ``A critique and improvement of the cl common
  language effect size statistics of mcgraw and wong,'' \emph{Journal of
  Educational and Behavioral Statistics}, vol.~25, no.~2, pp. 101--132, 2000.

\bibitem{Rojas:2015}
J.~M. Rojas, J.~Campos, M.~Vivanti, G.~Fraser, and A.~Arcuri, ``Combining
  multiple coverage criteria in search-based unit test generation,'' in
  \emph{Search-Based Software Engineering: 7th International Symposium}, 2015,
  pp. 93--108.

\bibitem{Kavir:2011}
K.~Shrestha and M.~J. Rutherford, ``An empirical evaluation of assertions as
  oracles,'' in \emph{2011 Fourth IEEE International Conference on Software
  Testing, Verification and Validation}, 2011, pp. 110--119.

\bibitem{Gunel:21}
G.~Jahangirova, D.~Clark, M.~Harman, and P.~Tonella, ``An empirical validation
  of oracle improvement,'' \emph{IEEE Transactions on Software Engineering},
  vol.~47, no.~8, pp. 1708--1728, 2021.

\bibitem{Valerio:21}
V.~Terragni, G.~Jahangirova, P.~Tonella, and M.~Pezzè, ``Gassert: A fully
  automated tool to improve assertion oracles,'' in \emph{2021 IEEE/ACM 43rd
  International Conference on Software Engineering: Companion Proceedings
  (ICSE-Companion)}, 2021, pp. 85--88.

\bibitem{Lan:19}
Z.~Lan, M.~Chen, S.~Goodman, K.~Gimpel, P.~Sharma, and R.~Soricut, ``Albert: A
  lite bert for self-supervised learning of language representations,''
  \emph{arXiv preprint arXiv:1909.11942}, 2019.

\bibitem{zhang:20}
Y.~Zhang, S.~Sun, M.~Galley, Y.-C. Chen, C.~Brockett, X.~Gao, J.~Gao, J.~Liu,
  and B.~Dolan, ``{DIALOGPT} : Large-scale generative pre-training for
  conversational response generation,'' in \emph{Proc. of ACL}, 2020.

\bibitem{Pilault:20}
J.~Pilault, R.~Li, S.~Subramanian, and C.~Pal, ``On extractive and abstractive
  neural document summarization with transformer language models,'' in
  \emph{Proc. of EMNLP}, 2020, pp. 9308--9319.

\bibitem{Cai:21}
X.~Cai, S.~Liu, J.~Han, L.~Yang, Z.~Liu, and T.~Liu, ``Chestxraybert: A
  pretrained language model for chest radiology report summarization,''
  \emph{IEEE Transactions on Multimedia}, pp. 845 -- 855, 2021.

\bibitem{Khashabi:20}
D.~Khashabi, S.~Min, T.~Khot, A.~Sabharwal, O.~Tafjord, P.~Clark, and
  H.~Hajishirzi, ``Unifiedqa: Crossing format boundaries with a single qa
  system,'' \emph{arXiv preprint arXiv:2005.00700}, 2020.

\bibitem{Cho:14}
K.~Cho, B.~Van~Merri{\"e}nboer, C.~Gulcehre, D.~Bahdanau, F.~Bougares,
  H.~Schwenk, and Y.~Bengio, ``Learning phrase representations using rnn
  encoder-decoder for statistical machine translation,'' \emph{arXiv preprint
  arXiv:1406.1078}, 2014.

\bibitem{Chen:21}
M.~Chen, J.~Tworek, H.~Jun, Q.~Yuan, H.~P. d.~O. Pinto, J.~Kaplan, H.~Edwards,
  Y.~Burda, N.~Joseph, G.~Brockman \emph{et~al.}, ``Evaluating large language
  models trained on code,'' \emph{arXiv preprint arXiv:2107.03374}, 2021.

\bibitem{Bui:21}
N.~D. Bui, Y.~Yu, and L.~Jiang, ``Infercode: Self-supervised learning of code
  representations by predicting subtrees,'' in \emph{Proc. of ICSE}.\hskip 1em
  plus 0.5em minus 0.4em\relax IEEE, 2021, pp. 1186--1197.

\bibitem{Peters:18}
M.~Peters, M.~Neumann, M.~Iyyer, M.~Gardner, C.~Clark, K.~Lee, and
  L.~Zettlemoyer, ``Deep contextualized word representations. arxiv 2018,''
  \emph{arXiv preprint arXiv:1802.05365}, vol.~12, 2018.

\bibitem{Copilot}
G.~Copilot, ``Your ai pair programmer,'' 2023,
  https://github.com/features/copilot/.

\bibitem{Miller:76}
W.~Miller and D.~L. Spooner, ``Automatic generation of floating-point test
  data,'' \emph{IEEE Transactions on Software Engineering}, no.~3, pp.
  223--226, 1976.

\bibitem{Li:07}
Z.~Li, M.~Harman, and R.~M. Hierons, ``Search algorithms for regression test
  case prioritization,'' \emph{IEEE Transactions on Software Engineering},
  vol.~33, no.~4, pp. 225--237, 2007.

\bibitem{Silva:17}
R.~A. Silva, S.~d. R.~S. de~Souza, and P.~S.~L. de~Souza, ``A systematic review
  on search based mutation testing,'' \emph{Information and Software
  Technology}, vol.~81, pp. 19--35, 2017.

\bibitem{Walcott:06}
K.~R. Walcott, M.~L. Soffa, G.~M. Kapfhammer, and R.~S. Roos, ``Time-aware test
  suite prioritization,'' in \emph{Proc. of ISSTA}, 2006, pp. 1--12.

\bibitem{Grano:19}
G.~Grano, C.~Laaber, A.~Panichella, and S.~Panichella, ``Testing with fewer
  resources: An adaptive approach to performance-aware test case generation,''
  \emph{IEEE Transactions on Software Engineering}, vol.~47, no.~11, pp.
  2332--2347, 2019.

\bibitem{Arcuri:21}
A.~Arcuri and J.~P. Galeotti, ``Enhancing search-based testing with testability
  transformations for existing apis,'' \emph{ACM Transactions on Software
  Engineering and Methodology}, vol.~31, no.~1, pp. 1--34, 2021.

\bibitem{Lin:20}
Y.~Lin, J.~Sun, G.~Fraser, Z.~Xiu, T.~Liu, and J.~S. Dong, ``Recovering fitness
  gradients for interprocedural boolean flags in search-based testing,'' in
  \emph{Proc. of ISSTA}, 2020, pp. 440--451.

\bibitem{Pietro:17}
P.~Braione, G.~Denaro, A.~Mattavelli, and M.~Pezz{\`e}, ``Combining symbolic
  execution and search-based testing for programs with complex heap inputs,''
  in \emph{Proc. of ISSTA}, 2017, pp. 90--101.

\bibitem{Xu:17}
X.~Xu, Z.~Zhu, and L.~Jiao, ``An adaptive fitness function based on branch
  hardness for search based testing,'' in \emph{Proc. of GECCO}, 2017, pp.
  1335--1342.

\bibitem{Miguel:15}
J.~M. Rojas, J.~Campos, M.~Vivanti, G.~Fraser, and A.~Arcuri, ``Combining
  multiple coverage criteria in search-based unit test generation,'' in
  \emph{Search-Based Software Engineering}, M.~Barros and Y.~Labiche, Eds.,
  2015, pp. 93--108.

\bibitem{Gay:17}
G.~Gay, ``Generating effective test suites by combining coverage criteria,'' in
  \emph{Search Based Software Engineering}, 2017, pp. 65--82.

\bibitem{Daka:17}
E.~Daka, J.~M. Rojas, and G.~Fraser, ``Generating unit tests with descriptive
  names or: Would you name your children thing1 and thing2?'' in \emph{Proc. of
  ISSTA}, 2017, pp. 57--67.

\bibitem{Roy:20}
D.~Roy, Z.~Zhang, M.~Ma, V.~Arnaoudova, A.~Panichella, S.~Panichella,
  D.~Gonzalez, and M.~Mirakhorli, ``Deeptc-enhancer: Improving the readability
  of automatically generated tests,'' in \emph{Proc. of ASE}, 2020, pp.
  287--298.

\bibitem{Wang:21}
S.~Wang, N.~Shrestha, A.~K. Subburaman, J.~Wang, M.~Wei, and N.~Nagappan,
  ``Automatic unit test generation for machine learning libraries: How far are
  we?'' in \emph{Proc. of ICSE}, 2021, pp. 1548--1560.

\bibitem{Dong:2020}
Z.~Dong, M.~B{\"o}hme, L.~Cojocaru, and A.~Roychoudhury, ``Time-travel testing
  of android apps,'' in \emph{Proc. of ICSE}, 2020, pp. 481--492.

\bibitem{Martin:2021}
A.~Martin-Lopez, S.~Segura, and A.~Ruiz-Cort{\'e}s, ``Restest: automated
  black-box testing of restful web apis,'' in \emph{Proc. of ISSTA}, 2021, pp.
  682--685.

\bibitem{Haq:2021}
F.~U. Haq, D.~Shin, L.~C. Briand, T.~Stifter, and J.~Wang, ``Automatic test
  suite generation for key-points detection dnns using many-objective search
  (experience paper),'' in \emph{Proc. of ISSTA}, 2021, pp. 91--102.

\end{thebibliography}


\balance
\end{document}